\documentclass[prl,aps,superscriptaddress,twocolumn]{revtex4-1}
\setcitestyle{super}
\usepackage{amsfonts,amssymb,amscd,amsthm}
\usepackage{graphicx}
\usepackage{mathrsfs}
\usepackage[intlimits]{amsmath}
\usepackage[colorlinks, citecolor=red]{hyperref}
\usepackage{lineno}
\usepackage{etoolbox}
\begin{document}

\title{Verifying quantum information scrambling dynamics in a fully controllable superconducting quantum simulator}

\author{J.-H. Wang}
\thanks{These two authors contributed equally to this work.}
\affiliation{Center for Quantum Information, Institute for Interdisciplinary Information Sciences, Tsinghua University, Beijing 100084, China}

\author{T.-Q. Cai}
\thanks{These two authors contributed equally to this work.}
\affiliation{Center for Quantum Information, Institute for Interdisciplinary Information Sciences, Tsinghua University, Beijing 100084, China}

\author{X.-Y. Han}
\affiliation{Center for Quantum Information, Institute for Interdisciplinary Information Sciences, Tsinghua University, Beijing 100084, China}

\author{Y.-W Ma}
\affiliation{Center for Quantum Information, Institute for Interdisciplinary Information Sciences, Tsinghua University, Beijing 100084, China}

\author{Z.-L Wang}
\affiliation{Center for Quantum Information, Institute for Interdisciplinary Information Sciences, Tsinghua University, Beijing 100084, China}

\author{Z.-H Bao}
\affiliation{Center for Quantum Information, Institute for Interdisciplinary Information Sciences, Tsinghua University, Beijing 100084, China}

\author{Y. Li}
\affiliation{Center for Quantum Information, Institute for Interdisciplinary Information Sciences, Tsinghua University, Beijing 100084, China}

\author{H.-Y Wang}
\affiliation{Center for Quantum Information, Institute for Interdisciplinary Information Sciences, Tsinghua University, Beijing 100084, China}

\author{H.-Y Zhang}
\affiliation{Center for Quantum Information, Institute for Interdisciplinary Information Sciences, Tsinghua University, Beijing 100084, China}

\author{L.-Y Sun}
\affiliation{Center for Quantum Information, Institute for Interdisciplinary Information Sciences, Tsinghua University, Beijing 100084, China}

\author{Y.-K. Wu}\email{wyukai@mail.tsinghua.edu.cn}
\affiliation{Center for Quantum Information, Institute for Interdisciplinary Information Sciences, Tsinghua University, Beijing 100084, China}

\author{Y.-P. Song}\email{ypsong@mail.tsinghua.edu.cn}
\affiliation{Center for Quantum Information, Institute for Interdisciplinary Information Sciences, Tsinghua University, Beijing 100084, China}

\author{L.-M. Duan}\email{lmduan@tsinghua.edu.cn}
\affiliation{Center for Quantum Information, Institute for Interdisciplinary Information Sciences, Tsinghua University, Beijing 100084, China}

\maketitle

\textbf{Quantum simulation elucidates properties of quantum many-body systems by mapping its Hamiltonian to a better-controlled system \cite{Feynman1982simulation,Lloyd1996universal,RevModPhys.86.153}. Being less stringent than a universal quantum computer, noisy small- and intermediate-scale quantum simulators have successfully demonstrated qualitative behavior such as phase transition, localization and thermalization \cite{houck2012chip,RevModPhys.93.025001,cold_atoms_review} which are insensitive to imperfections in the engineered Hamiltonian. For more complicated features like quantum information scrambling \cite{Hayden_2007,Sekino_2008}, higher controllability will be desired to simulate both the forward and the backward time evolutions and to diagnose experimental errors \cite{yoshida2017efficient,PhysRevX.9.011006}, which has only been achieved for discrete gates \cite{Landsman:2019aa,PhysRevX.11.021010}. Here, we study the verified scrambling in a 1D spin chain by an analogue superconducting quantum simulator with the signs and values of individual driving and coupling terms fully controllable. We measure the temporal and spatial patterns of out-of-time-ordered correlators \cite{kitaev2015otoc,swingle2018unscrambling} (OTOC) by engineering opposite Hamiltonians on two subsystems, with the Hamiltonian mismatch and the decoherence extracted quantitatively from the scrambling dynamics. Our work demonstrates the superconducting system as a powerful quantum simulator.}

Quantum information scrambling describes the spreading of information in a many-body quantum system \cite{Hayden_2007,Sekino_2008}. Being unseen to local observables, it is often probed by the OTOC \cite{kitaev2015otoc,swingle2018unscrambling}, namely, $C(t;W,V)\equiv \langle W^\dag(t) V^\dag(0) W(t) V(0)\rangle$ of two initially commuting operators $V(0)$ and $W(0)$, where $W(t)\equiv e^{i H t}W(0) e^{-i H t}$ is the time-evolved operator under the Hamiltonian $H$ of the system. As the information spreads, the OTOC decays with the increased nonlocality of $W(t)$. This information scrambling measured by the OTOC  is also a key concept in the fields of quantum chaos and quantum gravity \cite{Shenker2014,Maldacena2016}.

Despite the significant theoretical importance, OTOC is notoriously difficult to measure in experiments as it involves the correlation of operators at different time points. For a quantum simulator, this would require the capability to reverse each term in the Hamiltonian to obtain the forward and the backward propagation in time \cite{PhysRevA.94.040302}, which is previously mainly achieved for discrete gates at the cost of multiple layers \cite{PhysRevX.7.031011,mi2021information}, and for analogue simulation of only restricted classes of model Hamiltonians in atomic or NMR systems \cite{Garttner:2017aa,PhysRevLett.120.070501,PhysRevA.100.013623}. (An approximate scheme has also been proposed \cite{PhysRevX.9.021061} and demonstrated \cite{nie2019detecting,PhysRevLett.124.240505} recently using randomized measurements without the need of time reflection). Moreover, the measured decay of OTOC can come from both the quantum information scrambling and the experimental errors and noise such as the decoherence and the mismatch in the engineered Hamiltonian, making it a challenging task to verify the occurrence of scrambling. Recently it is proposed that by preparing two copies of the system in Einstein–Podolsky–Rosen (EPR) states and evolving them reversely in time under opposite Hamiltonians, scrambling and noise effects can be quantified separately by a teleportation-based scheme \cite{yoshida2017efficient,PhysRevX.9.011006}. This scheme has been tested by gate-based quantum scramblers \cite{Landsman:2019aa,PhysRevX.11.021010}, but its application in analogue quantum simulation and scrambling dynamics is still lacking due to the limited controllability of the simulated Hamiltonian.

\begin{figure*}[hbt]
\includegraphics{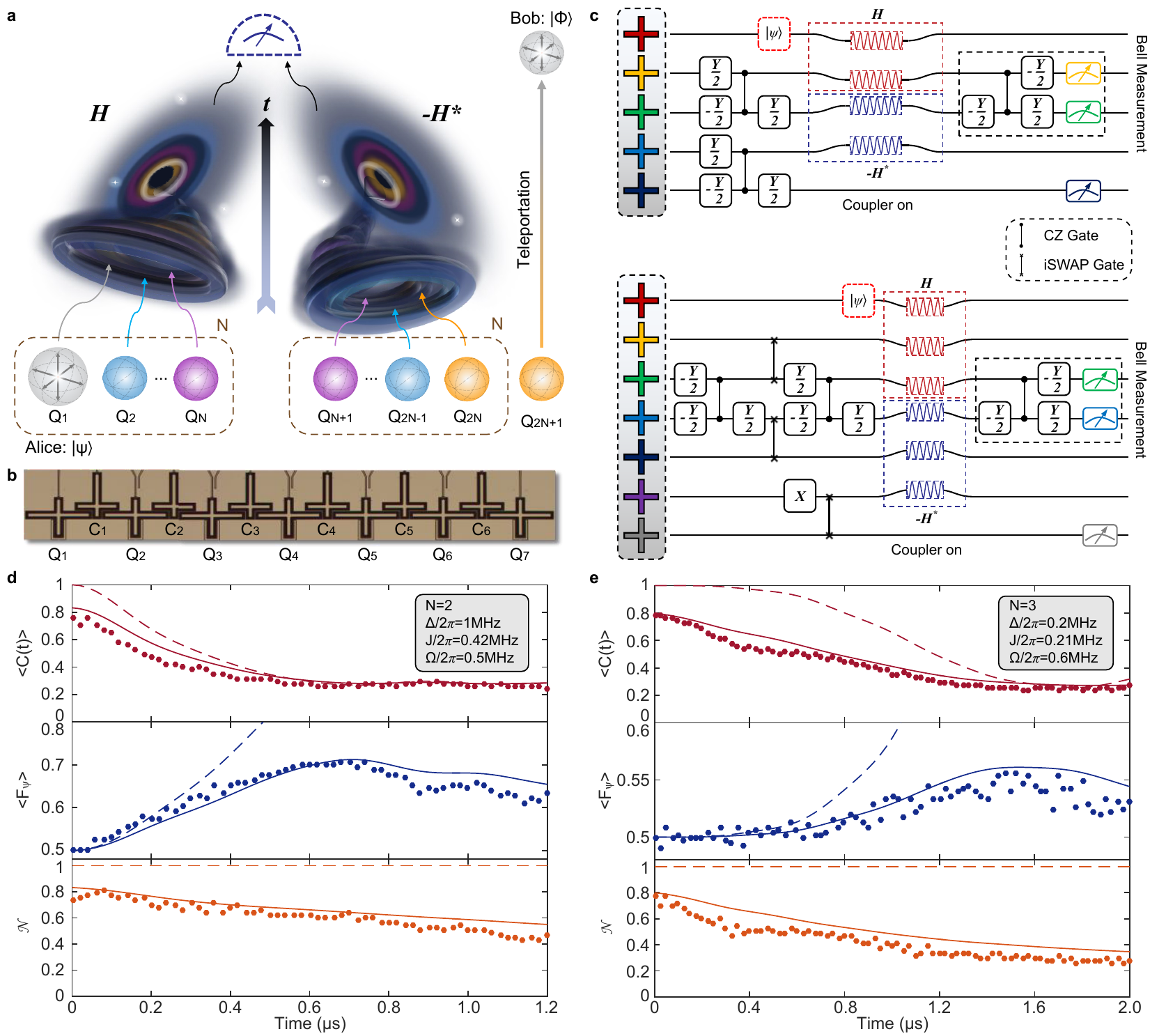}
\caption{\textbf{Experimental scheme and typical quantum simulation results.} \textbf{a}, Measuring quantum information scrambling dynamics using the teleportation-based protocol \cite{yoshida2017efficient,PhysRevX.9.011006}. Alice drops a secret state $|\psi \rangle$ ($Q_1$) into a black hole ($Q_2$-$Q_N$). Using only the Hawking radiation ($Q_N$) and an entangled system with the black hole ($Q_{N+1}$-$Q_{2N}$), Bob can decode the state ($Q_{2N+1}$), and the success of the teleportation verifies the scrambling. Qubit pairs in the same color represent an EPR state. \textbf{b}, Optical micrograph of the superconducting quantum processor with seven transmon qubits ($Q_1$-$Q_7$) and six couplers ($C_1$-$C_6$) used in this experiment. Crucially, the superconducting circuit utilizes the tunable coupler for realizing a unique competition between a positive direct and negative indirect coupling to achieve a continuous tunability, thus allowing a full control of qubit interactions. \textbf{c}, The schematic circuit using five qubits ($N=2$, upper) and seven qubits ($N=3$, lower). \textbf{d, e}, Typical experimental results (dots) under a general non-integrable Hamiltonian [Eq.~(\ref{eq:H})] for the average OTOC (upper), teleportation fidelity (middle) and noise parameter (lower, the deviation from one indicates experimental noise). \textbf{d} and \textbf{e} are for the five-qubit and seven-qubit cases, respectively. The dashed lines are the ideal results and the solid lines are the results in consideration of the imperfect EPR state preparation, Bell measurement and decoherence. The teleportation fidelity, reflecting absence of quantum scrambling at its minimal value $\langle F_{\psi} \rangle=0.5$, rises and reach a maximum but declines afterwards at a critical  evolution time where noise starts to break down the scrambling. The extracted noise parameter shows a constant decline over the course of the evolution, indicating the presence of decoherence error during the scrambling dynamics.}
\label{fig:Fig1}
\end{figure*}

In this work, we measure the OTOC evolution of a 1D spin chain Hamiltonian on a superconducting quantum simulator. Previously, superconducting qubits have been exploited to simulate various phenomena such as equilibrium and dynamical properties of spin chains and cavity QED systems \cite{houck2012chip,mit2020sc_review,huang2020superconducting}. With the help of tunable couplers \cite{PhysRevApplied.10.054062,PhysRevApplied.14.024070}, here we can precisely adjust the sign and the value of the coupling between each neighbouring pair of qubits, thus directly achieving opposite Hamiltonians on two subsystems. By initializing EPR states between the two subsystems and by performing Bell measurements after the simulated evolution, we measure the OTOC and the noise effects for various evolution time and verify whether the quantum scrambling occurs or not for different simulated Hamiltonians. Our work demonstrates the strong controllability of the superconducting quantum simulator, and can be applied to simulating diverse properties of complicated quantum many-body systems.

As shown in Fig.~\ref{fig:Fig1}\textbf{a}, our scheme to measure the OTOC utilizes the Hayden-Preskill variant of the black-hole information problem \cite{Hayden_2007,yoshida2017efficient,PhysRevX.9.011006}.
Suppose Alice drops a secret quantum state $|\psi\rangle$ ($Q_1$) into a black hole ($Q_2$-$Q_N$), which is in maximal entanglement with another system ($Q_{N+1}$-$Q_{2N-1}$) in the possession of Bob. Assuming full scrambling dynamics $U$ of the black hole, Bob can decode this secret state by collecting the Hawking radiation ($Q_N$) from the black hole, together with an auxiliary EPR state ($Q_{2N}$ and $Q_{2N+1}$). Specifically, a probabilistic decoder can be used \cite{yoshida2017efficient}, which evolves Bob's system ($Q_{N+1}$-$Q_{2N}$) reversely by $U^*$. Then upon projecting the Hawking radiation from the black hole ($Q_N$) and the counterpart of Bob's system ($Q_{N+1}$) into an EPR state, Bob can recover $|\psi\rangle$ and teleport it into a reference qubit ($Q_{2N+1}$). Now if instead of a black hole, the fastest information scrambler, we consider a many-body quantum system evolving under a Hamiltonian $H$ with the reverse evolution governed by $-H^*$, then this scheme allows probing the scrambling dynamics: The success rate of the projective measurement reveals the average OTOC, and the teleportation fidelity verifies true quantum information scrambling; combining the two results, one can further extract a noise parameter to characterize experimental imperfections in the scheme \cite{PhysRevX.9.011006}.

Our fully controllable superconducting quantum simulator consists of seven transmon qubits in 1D configuration, with each pair of neighbouring qubits mediated via a frequency-tunable coupler, as depicted in Fig.~\ref{fig:Fig1}\textbf{b} (see Supplementary Materials). Through the competition between a positive direct coupling and a negative indirect coupling, full control of qubit interactions can be achieved to engineer $H$ and $-H^*$ on the two subsystems. As shown in Fig.~\ref{fig:Fig1}\textbf{c}, we implement a five-qubit scheme ($N=2$) and a seven-qubit scheme ($N=3$) on this setup. We initialize the first qubit in $| \psi \rangle$ and the other qubits in EPR states using the quantum circuit consisting of single and two-qubit gates (see Supplementary Materials). Then we simulate Hamiltonian $H$ on $Q_1$-$Q_N$
\begin{equation}\label{eq:H}
H = \sum_{i=1}^{N} \left(\Delta_i \sigma_z^{i} + \Omega_i \sigma_x^{i}\right) + \sum_{i=1}^{N-1} J_{i,i+1} \sigma_z^{i} \sigma_z^{i+1}
\end{equation}
by applying microwave drive on individual qubits with amplitudes $\Omega_i$ and frequency detuning $\Delta_i$, and by coupling adjacent qubits with strength $J_{i,i+1}$ via the tunable couplers (see Supplementary Materials, where we also show results for simulating $\sigma_+^i\sigma_-^{i+1}+\sigma_-^i\sigma_+^{i+1}$ interactions). Similarly, we achieve $-H^*$ on $Q_{N+1}$-$Q_{2N}$. After a controllable evolution time $t$, we finally perform Bell measurement on $Q_N$ and $Q_{N+1}$ using another set of single-qubit and two-qubit gates. Ideally, the probability $P_{\psi}(t)$ to project into the EPR state $(|00\rangle+|11\rangle)/\sqrt{2}$ is related to the average OTOC $\langle C(t)\rangle$ by $\langle C(t)\rangle\equiv \iint d O_1 dO_N C(t;O_1,O_N)=\int d\psi P_{\psi}(t)$. In this equation, $O_1$ and $O_N$ are unitary operators acting on $Q_1$ and $Q_N$ respectively, and are averaged over the Haar measure; while the input state $|\psi\rangle$ can be averaged over a complex projective 1-design, say, $\{|0\rangle,|1\rangle\}$ \cite{PhysRevX.9.011006}. Furthermore, conditioned on the successful projective measurement, the teleportation fidelity $F_{\psi}\equiv \langle \psi | \rho_{2N+1} |\psi\rangle$ gives an additional characterization of the average OTOC. Here $\rho_{2N+1}$ is the final state of the reference qubit $Q_{2N+1}$ conditioned on the successful Bell measurement, and we have $\int d\psi P_{\psi} F_{\psi}=[\langle C(t)\rangle + 1/d]/(d+1)$, where $d=2$ is the dimension of the input state, and $|\psi\rangle$ is to be averaged over a complex projective 2-design, say, the six eigenstates of the Pauli operators $\sigma_x$, $\sigma_y$ and $\sigma_z$ \cite{PhysRevX.9.011006}. Finally, in the presence of decoherence and errors, $\langle \widetilde{C}(t)\rangle=\int d\psi P_{\psi}(t)$ will include the noise effects, and we can introduce a noise parameter $\mathcal{N}$ satisfying \cite{PhysRevX.9.011006,Landsman:2019aa} $\int d\psi P_{\psi} F_{\psi}=[\langle \widetilde{C}(t)\rangle + \mathcal{N}/d]/(d+1)$. $\mathcal{N}=1$ indicates the error-free case, while $\mathcal{N}<1$ reflects the experimental noise.

\begin{figure}[bt]
\includegraphics{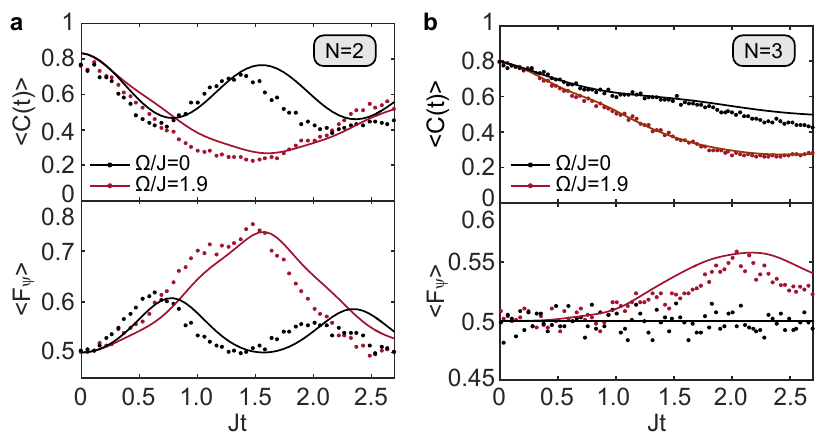}
\caption{\textbf{Scrambling dynamics for integrable Hamiltonian.} Upper and lower panels are the average OTOC and the teleportation fidelity under an integrable Hamiltonian ($\Delta_i=0$) with or without the $\sigma_x$ term (red and black, respectively). \textbf{a} and \textbf{b} are for the five-qubit and seven-qubit cases, respectively. The solid curves are the theoretical results with the SPAM errors and decoherence included. For a comparison between the two cases, the horizontal axes are scaled by $J$, the coupling in the Hamiltonian.}
\label{fig:Fig2}
\end{figure}

\begin{figure*}[bt]
\includegraphics{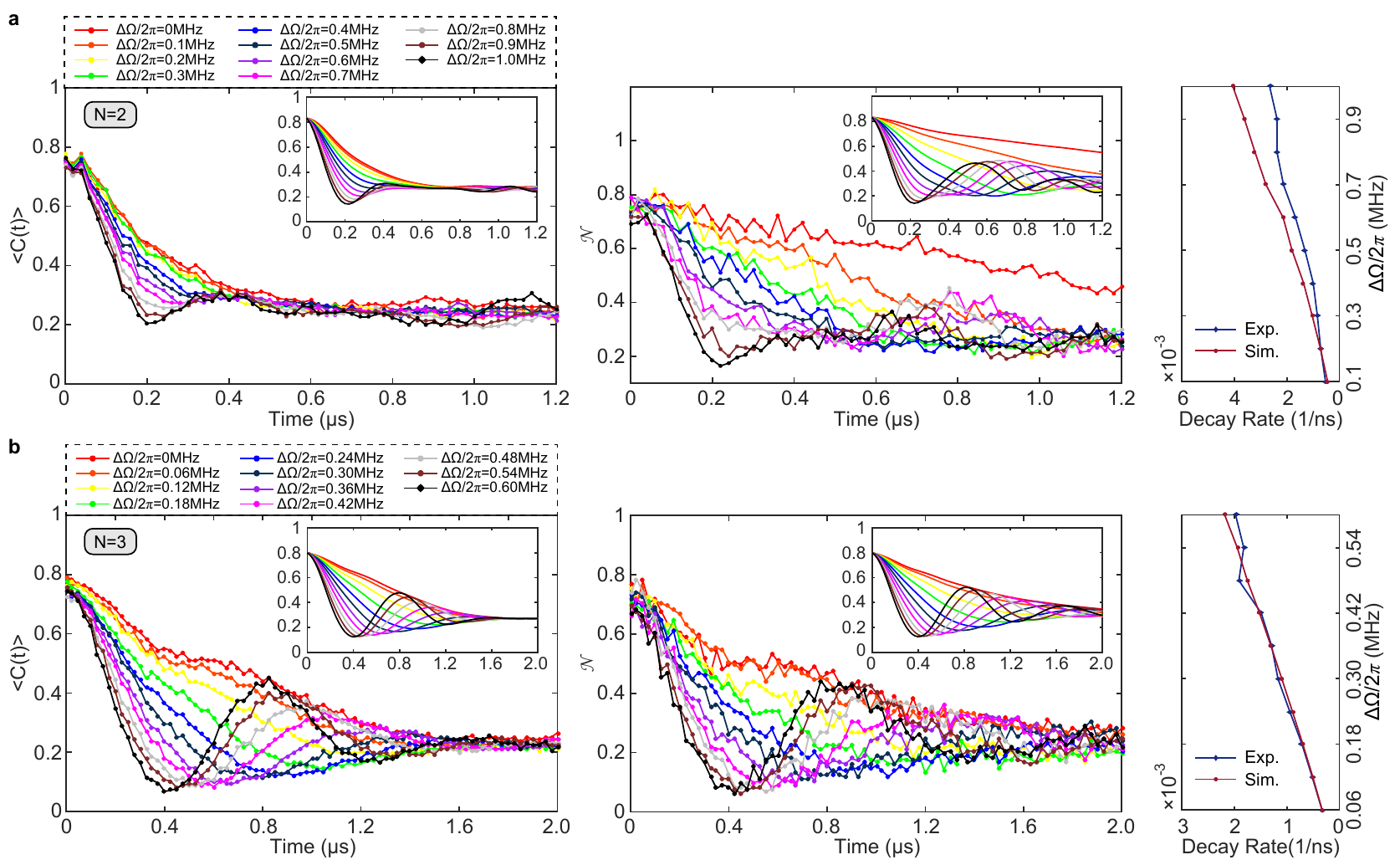}
\caption{\textbf{Effects of mismatch in the engineered opposite Hamiltonian.} \textbf{a} and \textbf{b} are for the five-qubit and seven-qubit cases, respectively. Left panels are the evolution of average OTOC under various mismatch strength $\Delta\Omega$ in the driving term, and the middle panels are those for the noise parameter. Insets are the theoretical results with the decoherence and SPAM errors included. As the mismatch increases, the noise parameter $\mathcal{N}$ decays faster from one, correctly indicating the error accumulation during the time evolution, while the starting point of the curves remains unchanged which is caused by the SPAM error. Right panels are the decay rates of $\mathcal{N}(t)$ versus $\Delta\Omega$. We extract the decay rate by linear fitting to the initial part of the curves before $\mathcal{N}(t)$ decreases to $0.4$, showing a near-linear growth with the mismatch.}
\label{fig:Fig3}
\end{figure*}

Typical experimental results under the general non-integrable Hamiltonian [Eq.~(\ref{eq:H})] are presented in Fig.~\ref{fig:Fig1}\textbf{d, e} for the $N=2$ and $N=3$ cases. The average OTOC decays with time, which comes both from information propagation and from the experimental noise and errors. On the other hand, the teleportation fidelity starts from 0.5 for a fully random state, rises to a maximum value verifying the occurrence of information scrambling, and then decays again due to the noise effects. Also note that the fidelity increases slower for $N=3$ than for $N=2$ since it takes longer time for the information to propagate.
Finally, the noise parameter has an initial deviation from one because of the state-preparation-and-measurement (SPAM) errors, and then decays further due to the decoherence and the Hamiltonian mismatch. Note that the experimental results (dots) agree well with the theoretical results (solid lines) considering the SPAM error and the coherence time of the qubits. Therefore the mismatch in the engineered Hamiltonian only contributes small effects to the deviation from the ideal evolution (dashed lines), demonstrating the accurate controllability of our tunable quantum simulator.

By tuning the parameters of the Hamiltonian, we can also simulate integrable models with distinct dynamics from the non-integrable one. In Fig.~\ref{fig:Fig2} we set the driving pulses in resonance with each qubit. Then we have $\Delta_i=0$ and the Hamiltonian reduces to a 1D transverse-field Ising model (red). In this situation the OTOC and the teleportation fidelity oscillate, indicating the information bouncing back and forth in the system. We can further turn off the driving pulse so that $\Omega_i=0$ (black). Then the remaining terms in the Hamiltonian $\sigma_z^i\sigma_z^{i+1}$ commute with each other and the information does not propagate. As we can see, only two directly coupled qubits ($N=2$) show the oscillation, while for $N=3$ the teleportation fidelity stays around $\langle F_{\psi} \rangle=0.5$ (the average OTOC still decays due to the experimental decoherence).

As mentioned before, this scheme of measuring OTOC can diagnose not only the above incoherent errors but also the coherent ones in the experiment. As illustrated in Fig.~\ref{fig:Fig3}, we deliberatly introduce a mismatch in the two Hamiltonians $H$ and $-H^*$ by applying different driving strength $\Delta\Omega$. Clearly, the noise parameter $\mathcal{N}$ decays faster with increasing $\Delta\Omega$ and therefore reflects the coherent error in the time evolution. While at relatively small mismatches, the noise parameter $\mathcal{N}$ constantly decreases with elongating the evolution time, revealing that the decoherence error plays an important role over the course of scrambling dynamics. Also note that larger $\Delta\Omega$ leads to faster decay in the average OTOC from the ideal case (left panels), thus the necessity of using $\mathcal{N}$ to bound this effect \cite{PhysRevX.9.011006}.

\begin{figure}[bt]
\includegraphics{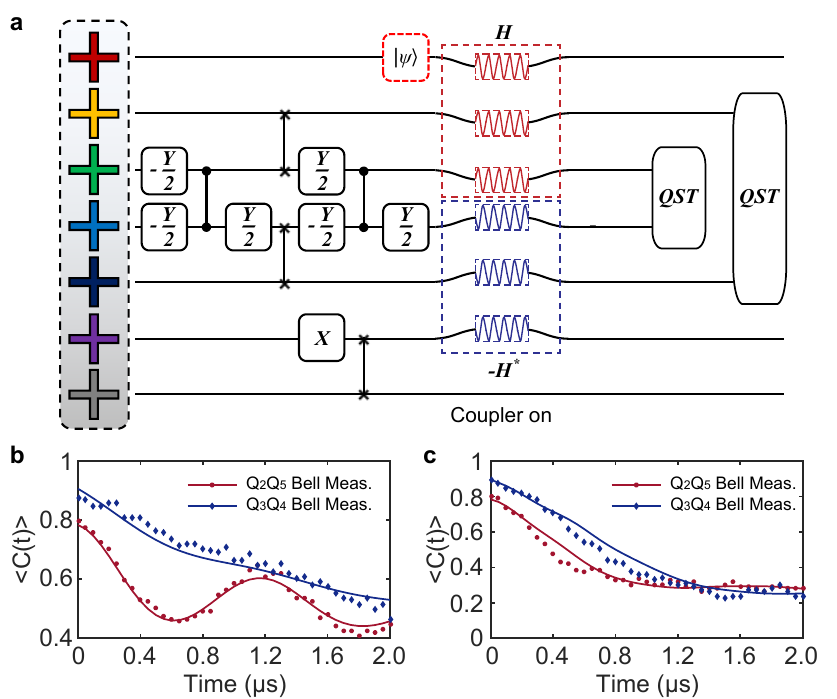}
\caption{\textbf{Average OTOC for qubits at different distances.} \textbf{a}, The schematic circuit. Here Bell measurements for distant qubit pairs are needed. To reduce the error due to imperfect two-qubit gates, we measure the two-qubit final states through quantum state tomography rather than swapping them into nearest neighbors. Then the Bell measurement success rate can be extracted. \textbf{b}, The average OTOC between sites 1 and 2 (red) and between sites 1 and 3 (blue) under a simulated Hamiltonian with $\Omega_i=0$. The solid curves are the theoretical results with the SPAM errors and decoherence included. \textbf{c}, Similar plot for a Hamiltonian with $\Omega_i \ne 0$. The OTOC drops faster for the adjacent qubit pair, which is a consequence of the finite information spreading speed.}
\label{fig:Fig4}
\end{figure}

Finally, we use the circuit in Fig.~\ref{fig:Fig4} to examine the spatially resolved information propagation in an $N=3$ model. By preparing $|\psi\rangle$ in $Q_1$ and performing Bell measurement on $Q_3$ and $Q_4$, we obtain average OTOC between sites 1 and 3. Now if we perform Bell measurement instead on $Q_2$ and $Q_5$, we will get average OTOC between sites 1 and 2. For this measurement, in principle we can first swap the qubit states of $Q_2$ and $Q_5$ into $Q_3$ and $Q_4$, and then make the projective measurement as before. However, this would introduce additional experimental errors due to the imperfect swapping operations. Therefore, here we directly perform quantum state tomography \cite{nielsen2000quantum} to reconstruct the desired two-qubit states, and then compute the success probability to project into an EPR state. In Fig.~\ref{fig:Fig4}\textbf{b} and \textbf{c}, we plot the results for an integrable model ($\Omega_i=0$) and a non-integrable model ($\Omega_i\ne 0$), respectively. In both cases, we see that the OTOC between the adjacent sites 1 and 2 decays faster than that between the distant sites 1 and 3, accordant with the finite information propagation speed on the 1D chain. We also observe that the initial OTOCs for the two pairs are different. This can be explained by the lower state preparation fidelity between $Q_2$ and $Q_5$ since they are swapped from $Q_3$ and $Q_4$ by additional two-qubit gates.

Our demonstration of teleportation-based OTOC measurement for unitary time evolution, instead of the gate-implemented digital scramblers \cite{Landsman:2019aa,PhysRevX.11.021010}, provides a powerful benchmarking tool for scrambling dynamics of many-body systems, and allows efficient noise diagnosis for future large-scale quantum processors. More complicated many-body models can be simulated by constructing larger system sizes and suppressing experimental errors with more sophisticated control techniques. Furthermore, by exploring different forms of system Hamiltonians and decoherence, the crossover from classical to quantum chaos can be examined \cite{PhysRevX.9.011006}. Therefore, our work shows the powerful controllability of the coupler-mediated superconducting system and opens the door toward more advanced applications of analogue quantum simulation in the future.

\makeatletter
\makeatother

\begin{thebibliography}{10}
\expandafter\ifx\csname url\endcsname\relax
  \def\url#1{\texttt{#1}}\fi
\expandafter\ifx\csname urlprefix\endcsname\relax\def\urlprefix{URL }\fi
\providecommand{\bibinfo}[2]{#2}
\providecommand{\eprint}[2][]{\url{#2}}

\bibitem{Feynman1982simulation}
\bibinfo{author}{Feynman, R.~P.}
\newblock \bibinfo{title}{Simulating physics with computers}.
\newblock \emph{\bibinfo{journal}{International Journal of Theoretical
  Physics}} \textbf{\bibinfo{volume}{21}}, \bibinfo{pages}{467--488}
  (\bibinfo{year}{1982}).

\bibitem{Lloyd1996universal}
\bibinfo{author}{Lloyd, S.}
\newblock \bibinfo{title}{Universal quantum simulators}.
\newblock \emph{\bibinfo{journal}{Science}} \textbf{\bibinfo{volume}{273}},
  \bibinfo{pages}{1073--1078} (\bibinfo{year}{1996}).

\bibitem{RevModPhys.86.153}
\bibinfo{author}{Georgescu, I.~M.}, \bibinfo{author}{Ashhab, S.} \&
  \bibinfo{author}{Nori, F.}
\newblock \bibinfo{title}{Quantum simulation}.
\newblock \emph{\bibinfo{journal}{Rev. Mod. Phys.}}
  \textbf{\bibinfo{volume}{86}}, \bibinfo{pages}{153--185}
  (\bibinfo{year}{2014}).

\bibitem{houck2012chip}
\bibinfo{author}{Houck, A.~A.}, \bibinfo{author}{T{\"u}reci, H.~E.} \&
  \bibinfo{author}{Koch, J.}
\newblock \bibinfo{title}{On-chip quantum simulation with superconducting
  circuits}.
\newblock \emph{\bibinfo{journal}{Nature Physics}}
  \textbf{\bibinfo{volume}{8}}, \bibinfo{pages}{292--299}
  (\bibinfo{year}{2012}).

\bibitem{RevModPhys.93.025001}
\bibinfo{author}{Monroe, C.} \emph{et~al.}
\newblock \bibinfo{title}{Programmable quantum simulations of spin systems with
  trapped ions}.
\newblock \emph{\bibinfo{journal}{Rev. Mod. Phys.}}
  \textbf{\bibinfo{volume}{93}}, \bibinfo{pages}{025001}
  (\bibinfo{year}{2021}).

\bibitem{cold_atoms_review}
\bibinfo{author}{Gross, C.} \& \bibinfo{author}{Bloch, I.}
\newblock \bibinfo{title}{Quantum simulations with ultracold atoms in optical
  lattices}.
\newblock \emph{\bibinfo{journal}{Science}} \textbf{\bibinfo{volume}{357}},
  \bibinfo{pages}{995--1001} (\bibinfo{year}{2017}).

\bibitem{Hayden_2007}
\bibinfo{author}{Hayden, P.} \& \bibinfo{author}{Preskill, J.}
\newblock \bibinfo{title}{Black holes as mirrors: quantum information in random
  subsystems}.
\newblock \emph{\bibinfo{journal}{Journal of High Energy Physics}}
  \textbf{\bibinfo{volume}{2007}}, \bibinfo{pages}{120--120}
  (\bibinfo{year}{2007}).

\bibitem{Sekino_2008}
\bibinfo{author}{Sekino, Y.} \& \bibinfo{author}{Susskind, L.}
\newblock \bibinfo{title}{Fast scramblers}.
\newblock \emph{\bibinfo{journal}{Journal of High Energy Physics}}
  \textbf{\bibinfo{volume}{2008}}, \bibinfo{pages}{065--065}
  (\bibinfo{year}{2008}).

\bibitem{yoshida2017efficient}
\bibinfo{author}{Yoshida, B.} \& \bibinfo{author}{Kitaev, A.}
\newblock \bibinfo{title}{Efficient decoding for the {Hayden}-{Preskill}
  protocol} (\bibinfo{year}{2017}).
\newblock arXiv:\eprint{1710.03363}.

\bibitem{PhysRevX.9.011006}
\bibinfo{author}{Yoshida, B.} \& \bibinfo{author}{Yao, N.~Y.}
\newblock \bibinfo{title}{Disentangling scrambling and decoherence via quantum
  teleportation}.
\newblock \emph{\bibinfo{journal}{Phys. Rev. X}} \textbf{\bibinfo{volume}{9}},
  \bibinfo{pages}{011006} (\bibinfo{year}{2019}).

\bibitem{Landsman:2019aa}
\bibinfo{author}{Landsman, K.~A.} \emph{et~al.}
\newblock \bibinfo{title}{Verified quantum information scrambling}.
\newblock \emph{\bibinfo{journal}{Nature}} \textbf{\bibinfo{volume}{567}},
  \bibinfo{pages}{61--65} (\bibinfo{year}{2019}).

\bibitem{PhysRevX.11.021010}
\bibinfo{author}{Blok, M.~S.} \emph{et~al.}
\newblock \bibinfo{title}{Quantum information scrambling on a superconducting
  qutrit processor}.
\newblock \emph{\bibinfo{journal}{Phys. Rev. X}} \textbf{\bibinfo{volume}{11}},
  \bibinfo{pages}{021010} (\bibinfo{year}{2021}).

\bibitem{kitaev2015otoc}
\bibinfo{author}{Kitaev, A.}
\newblock \bibinfo{title}{A simple model of quantum holography}.
\newblock
  \bibinfo{howpublished}{http://online.kitp.ucsb.edu/online/entangled15/kitaev/,
  http://online.kitp.ucsb.edu/online/entangled15/kitaev2/}
  (\bibinfo{year}{2015}).

\bibitem{swingle2018unscrambling}
\bibinfo{author}{Swingle, B.}
\newblock \bibinfo{title}{Unscrambling the physics of out-of-time-order
  correlators}.
\newblock \emph{\bibinfo{journal}{Nature Physics}}
  \textbf{\bibinfo{volume}{14}}, \bibinfo{pages}{988} (\bibinfo{year}{2018}).

\bibitem{Shenker2014}
\bibinfo{author}{Shenker, S.~H.} \& \bibinfo{author}{Stanford, D.}
\newblock \bibinfo{title}{Black holes and the butterfly effect}.
\newblock \emph{\bibinfo{journal}{Journal of High Energy Physics}}
  \textbf{\bibinfo{volume}{2014}}, \bibinfo{pages}{67} (\bibinfo{year}{2014}).

\bibitem{Maldacena2016}
\bibinfo{author}{Maldacena, J.}, \bibinfo{author}{Shenker, S.~H.} \&
  \bibinfo{author}{Stanford, D.}
\newblock \bibinfo{title}{A bound on chaos}.
\newblock \emph{\bibinfo{journal}{Journal of High Energy Physics}}
  \textbf{\bibinfo{volume}{2016}}, \bibinfo{pages}{106} (\bibinfo{year}{2016}).

\bibitem{PhysRevA.94.040302}
\bibinfo{author}{Swingle, B.}, \bibinfo{author}{Bentsen, G.},
  \bibinfo{author}{Schleier-Smith, M.} \& \bibinfo{author}{Hayden, P.}
\newblock \bibinfo{title}{Measuring the scrambling of quantum information}.
\newblock \emph{\bibinfo{journal}{Phys. Rev. A}} \textbf{\bibinfo{volume}{94}},
  \bibinfo{pages}{040302(R)} (\bibinfo{year}{2016}).

\bibitem{PhysRevX.7.031011}
\bibinfo{author}{Li, J.} \emph{et~al.}
\newblock \bibinfo{title}{Measuring out-of-time-order correlators on a nuclear
  magnetic resonance quantum simulator}.
\newblock \emph{\bibinfo{journal}{Phys. Rev. X}} \textbf{\bibinfo{volume}{7}},
  \bibinfo{pages}{031011} (\bibinfo{year}{2017}).

\bibitem{mi2021information}
\bibinfo{author}{Mi, X.} \emph{et~al.}
\newblock \bibinfo{title}{Information scrambling in quantum circuits}.
\newblock \emph{\bibinfo{journal}{Science}} \bibinfo{pages}{eabg5029}
  (\bibinfo{year}{2021}).

\bibitem{Garttner:2017aa}
\bibinfo{author}{G{\"a}rttner, M.} \emph{et~al.}
\newblock \bibinfo{title}{Measuring out-of-time-order correlations and multiple
  quantum spectra in a trapped-ion quantum magnet}.
\newblock \emph{\bibinfo{journal}{Nature Physics}}
  \textbf{\bibinfo{volume}{13}}, \bibinfo{pages}{781--786}
  (\bibinfo{year}{2017}).

\bibitem{PhysRevLett.120.070501}
\bibinfo{author}{Wei, K.~X.}, \bibinfo{author}{Ramanathan, C.} \&
  \bibinfo{author}{Cappellaro, P.}
\newblock \bibinfo{title}{Exploring localization in nuclear spin chains}.
\newblock \emph{\bibinfo{journal}{Phys. Rev. Lett.}}
  \textbf{\bibinfo{volume}{120}}, \bibinfo{pages}{070501}
  (\bibinfo{year}{2018}).

\bibitem{PhysRevA.100.013623}
\bibinfo{author}{Meier, E.~J.}, \bibinfo{author}{Ang'ong'a, J.},
  \bibinfo{author}{An, F.~A.} \& \bibinfo{author}{Gadway, B.}
\newblock \bibinfo{title}{Exploring quantum signatures of chaos on a {Floquet}
  synthetic lattice}.
\newblock \emph{\bibinfo{journal}{Phys. Rev. A}}
  \textbf{\bibinfo{volume}{100}}, \bibinfo{pages}{013623}
  (\bibinfo{year}{2019}).

\bibitem{PhysRevX.9.021061}
\bibinfo{author}{Vermersch, B.}, \bibinfo{author}{Elben, A.},
  \bibinfo{author}{Sieberer, L.~M.}, \bibinfo{author}{Yao, N.~Y.} \&
  \bibinfo{author}{Zoller, P.}
\newblock \bibinfo{title}{Probing scrambling using statistical correlations
  between randomized measurements}.
\newblock \emph{\bibinfo{journal}{Phys. Rev. X}} \textbf{\bibinfo{volume}{9}},
  \bibinfo{pages}{021061} (\bibinfo{year}{2019}).

\bibitem{nie2019detecting}
\bibinfo{author}{Nie, X.} \emph{et~al.}
\newblock \bibinfo{title}{Detecting scrambling via statistical correlations
  between randomized measurements on an {NMR} quantum simulator}
  (\bibinfo{year}{2019}).
\newblock arXiv:\eprint{1903.12237}.

\bibitem{PhysRevLett.124.240505}
\bibinfo{author}{Joshi, M.~K.} \emph{et~al.}
\newblock \bibinfo{title}{Quantum information scrambling in a trapped-ion
  quantum simulator with tunable range interactions}.
\newblock \emph{\bibinfo{journal}{Phys. Rev. Lett.}}
  \textbf{\bibinfo{volume}{124}}, \bibinfo{pages}{240505}
  (\bibinfo{year}{2020}).

\bibitem{mit2020sc_review}
\bibinfo{author}{Kjaergaard, M.} \emph{et~al.}
\newblock \bibinfo{title}{Superconducting qubits: Current state of play}.
\newblock \emph{\bibinfo{journal}{Annual Review of Condensed Matter Physics}}
  \textbf{\bibinfo{volume}{11}}, \bibinfo{pages}{369--395}
  (\bibinfo{year}{2020}).

\bibitem{huang2020superconducting}
\bibinfo{author}{Huang, H.-L.}, \bibinfo{author}{Wu, D.}, \bibinfo{author}{Fan,
  D.} \& \bibinfo{author}{Zhu, X.}
\newblock \bibinfo{title}{Superconducting quantum computing: a review}.
\newblock \emph{\bibinfo{journal}{Science China Information Sciences}}
  \textbf{\bibinfo{volume}{63}}, \bibinfo{pages}{180501}
  (\bibinfo{year}{2020}).

\bibitem{PhysRevApplied.10.054062}
\bibinfo{author}{Yan, F.} \emph{et~al.}
\newblock \bibinfo{title}{Tunable coupling scheme for implementing
  high-fidelity two-qubit gates}.
\newblock \emph{\bibinfo{journal}{Phys. Rev. Applied}}
  \textbf{\bibinfo{volume}{10}}, \bibinfo{pages}{054062}
  (\bibinfo{year}{2018}).

\bibitem{PhysRevApplied.14.024070}
\bibinfo{author}{Li, X.} \emph{et~al.}
\newblock \bibinfo{title}{Tunable coupler for realizing a controlled-phase gate
  with dynamically decoupled regime in a superconducting circuit}.
\newblock \emph{\bibinfo{journal}{Phys. Rev. Applied}}
  \textbf{\bibinfo{volume}{14}}, \bibinfo{pages}{024070}
  (\bibinfo{year}{2020}).

\bibitem{nielsen2000quantum}
\bibinfo{author}{Nielsen, M.} \& \bibinfo{author}{Chuang, I.}
\newblock \emph{\bibinfo{title}{Quantum Computation and Quantum Information}}
  (\bibinfo{publisher}{Cambridge University Press},
  \bibinfo{address}{Cambridge}, \bibinfo{year}{2010}), \bibinfo{edition}{10th
  anniversary} edn.
\end{thebibliography}

\bigskip

\textbf{Acknowledgements:} This work is supported by National Natural Science Foundation of China under Grant No.11874235 and 11674060, and the Tsinghua University Initiative Scientific Research Program. Y.K.W. acknowledges support from the start-up fund from Tsinghua University.

\end{document}


\title{Supplementary Materials for ``Verifying quantum information scrambling dynamics in a fully controllable superconducting quantum simulator''}

\author{J.-H. Wang}
\thanks{These two authors contributed equally to this work.}
\affiliation{Center for Quantum Information, Institute for Interdisciplinary Information Sciences, Tsinghua University, Beijing 100084, China}

\author{T.-Q. Cai}
\thanks{These two authors contributed equally to this work.}
\affiliation{Center for Quantum Information, Institute for Interdisciplinary Information Sciences, Tsinghua University, Beijing 100084, China}

\author{X.-Y. Han}
\affiliation{Center for Quantum Information, Institute for Interdisciplinary Information Sciences, Tsinghua University, Beijing 100084, China}

\author{Y.-W Ma}
\affiliation{Center for Quantum Information, Institute for Interdisciplinary Information Sciences, Tsinghua University, Beijing 100084, China}

\author{Z.-L Wang}
\affiliation{Center for Quantum Information, Institute for Interdisciplinary Information Sciences, Tsinghua University, Beijing 100084, China}

\author{Z.-H Bao}
\affiliation{Center for Quantum Information, Institute for Interdisciplinary Information Sciences, Tsinghua University, Beijing 100084, China}

\author{Y. Li}
\affiliation{Center for Quantum Information, Institute for Interdisciplinary Information Sciences, Tsinghua University, Beijing 100084, China}

\author{H.-Y Wang}
\affiliation{Center for Quantum Information, Institute for Interdisciplinary Information Sciences, Tsinghua University, Beijing 100084, China}

\author{H.-Y Zhang}
\affiliation{Center for Quantum Information, Institute for Interdisciplinary Information Sciences, Tsinghua University, Beijing 100084, China}

\author{L.-Y Sun}
\affiliation{Center for Quantum Information, Institute for Interdisciplinary Information Sciences, Tsinghua University, Beijing 100084, China}

\author{Y.-K. Wu}\email{wyukai@mail.tsinghua.edu.cn}
\affiliation{Center for Quantum Information, Institute for Interdisciplinary Information Sciences, Tsinghua University, Beijing 100084, China}

\author{Y.-P. Song}\email{ypsong@mail.tsinghua.edu.cn}
\affiliation{Center for Quantum Information, Institute for Interdisciplinary Information Sciences, Tsinghua University, Beijing 100084, China}

\author{L.-M. Duan}\email{lmduan@tsinghua.edu.cn}
\affiliation{Center for Quantum Information, Institute for Interdisciplinary Information Sciences, Tsinghua University, Beijing 100084, China}

\maketitle

\tableofcontents

\newpage

\section{Extended Experimental Data}

In our experiment, we mainly study the dynamics of the Hamiltonian in two scenarios: $ZZ$-type coupling scheme (integrable and nonintegrable) and $(XX+YY)$-type coupling scheme (integrable) which can be both naturally generated in our superconducting quantum processor. $ZZ$-type coupling represents the interaction of $\sigma_z^i \sigma_z^{i+1}$ in the Hamiltonian while $(XX+YY)$-type coupling defines the interaction as $\sigma_+^i \sigma_-^{i+1} + \sigma_-^i \sigma_+^{i+1}$ \cite{zhao2021suppression,stehlik2021tunable,ku2020suppression}. In our main text, we present the main work with the $ZZ$-type coupling scheme. In this section, we provide some extended data to further elaborate our results, including the experimental parameters for the ZZ-type coupling scheme, the OTOC dynamics and the effect of Hamiltonian mismatch in $(XX+YY)$-type coupling scheme.

\subsection{OTOC dynamics in $ZZ$-type coupling scheme}

\begin{table}[hbt]
\caption{Hamiltonian parameters in $ZZ$-type coupling scheme. $\Delta_{i}$ is the frequency detuning between the drive frequency and the frequency of each qubit. $\Omega_{i}$ is the drive pulse amplitude applied on each qubit while $J_{i,i+1}$ represents the $ZZ$ interaction between the nearest neighbor qubits in the subsystem in the OTOC dynamics.}
\begin{threeparttable}
\begin{tabular}{cp{0.9cm}<{\centering}p{0.75cm}<{\centering}p{0.75cm}<{\centering}p{0.75cm}<{\centering}p{0.75cm}<{\centering}p{0.75cm}<{\centering}p{0.75cm}<{\centering}p{0.75cm}<{\centering}}
\\[-2pt]
&\multicolumn{5}{c}{Five Qubit OTOC Experiment} \tabularnewline
\\[-7pt]
\hline
\hline
&{$Q_1$} &{$Q_2$} &{$Q_3$} &{$Q_4$} &{$Q_5$} &{$Q_6$} &{$Q_7$} \tabularnewline
\hline
$\Delta_{i}/2\pi$ (MHz) &{$1$} &{$1$} &{$-1$} &{$-1$} &{$0$} &{$\sim$} &{$\sim$} \tabularnewline
$\Omega_{i}/2\pi$ (MHz) &{$0.5$} &{$0.5$} &{$-0.5$} &{$-0.5$} &{$0$} &{$\sim$} &{$\sim$} \tabularnewline
\hline
\hline
&{$Q_{1-2}$} &{$Q_{2-3}$} &{$Q_{3-4}$} &{$Q_{4-5}$} &{$Q_{5-6}$} &{$Q_{6-7}$} \tabularnewline
\hline
$J_{i,i+1}/2\pi$ (MHz) &{$0.42$} &{$\approx 0$} &{$-0.42$} &{$\approx 0$} &{$\sim$} &{$\sim$}  \tabularnewline
\hline

\\[-5pt]
&\multicolumn{5}{c}{Seven Qubit OTOC Experiment} \tabularnewline
\\[-7pt]
\hline
\hline
&{$Q_1$} &{$Q_2$} &{$Q_3$} &{$Q_4$} &{$Q_5$} &{$Q_6$} &{$Q_7$} \tabularnewline
\hline
$\Delta_{i}/2\pi$ (MHz) &{$0,2$} &{$0.2$} &{$0.2$} &{$-0.2$} &{$-0.2$} &{$-0.2$} &{$0$} \tabularnewline
$\Omega_{i}/2\pi$ (MHz) &{$0,6$} &{$0.6$} &{$0.6$} &{$-0.6$} &{$-0.6$} &{$-0.6$} &{$0$} \tabularnewline
\hline
\hline
&{$Q_{1-2}$} &{$Q_{2-3}$} &{$Q_{3-4}$} &{$Q_{4-5}$} &{$Q_{5-6}$} &{$Q_{6-7}$} \tabularnewline
\hline
$J_{i,i+1}/2\pi$ (MHz) &{$0.21$} &{$0.21$} &{$\approx 0$} &{$-0.21$} &{$-0.21$} &{$\approx 0$}  \tabularnewline
\hline
\end{tabular} \vspace{0pt}
\label{Table:TableS1}
\end{threeparttable}
\end{table}

\begin{figure*}[hbt]
\includegraphics{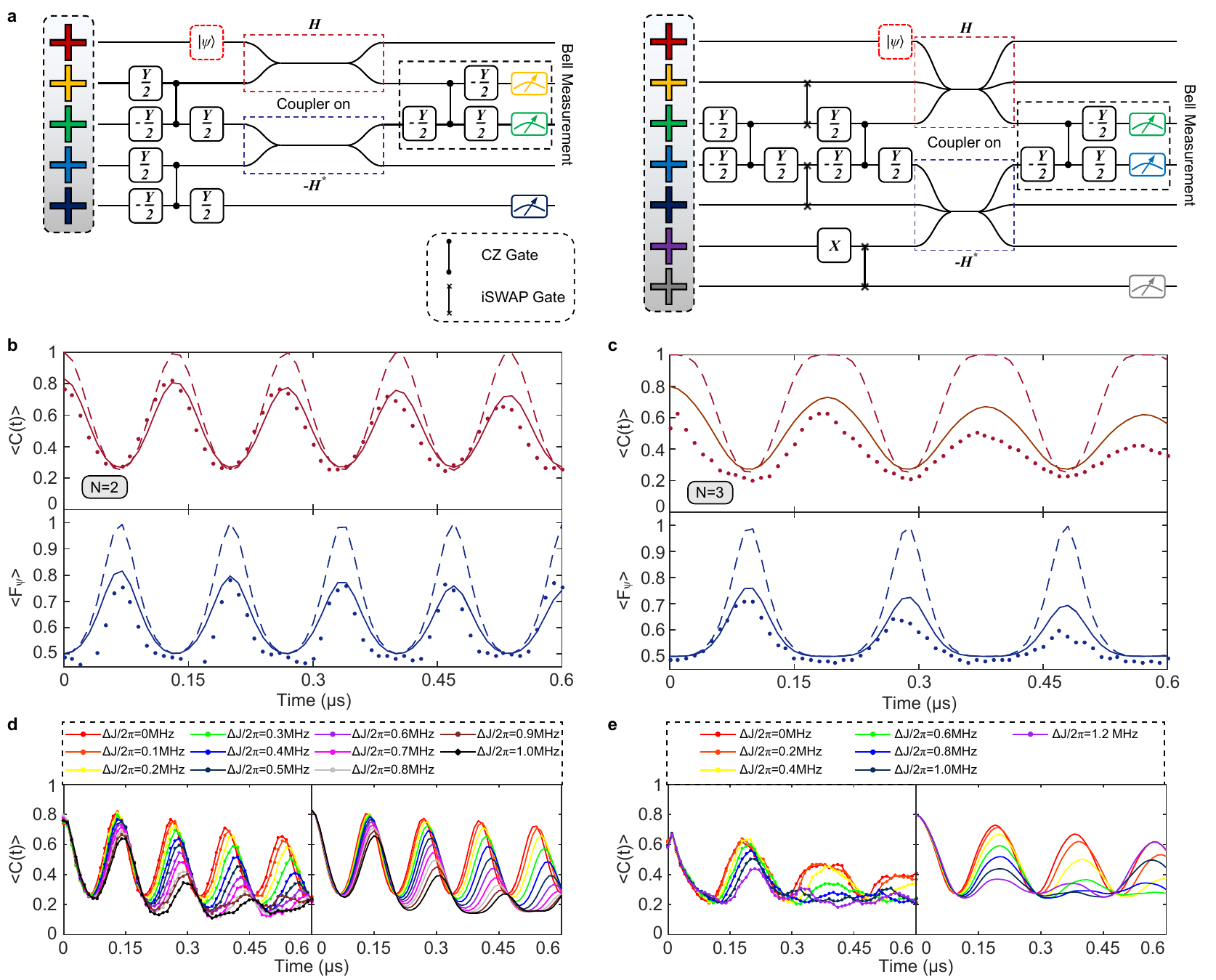}
\caption{\textbf{OTOC dynamics in $(XX+YY)$-type coupling scheme.} 
\textbf{a}, The schematic circuit using five qubits ($N=2$, left) and seven qubits ($N=3$, right) in the $(XX+YY)$-type coupling scheme.  \textbf{b, c}, Typical experimental results (dots) under the integrable Hamiltonian [Eq.~(\ref{Eq1-3})] for the average OTOC (upper) and teleportation fidelity (lower). \textbf{b} and \textbf{c} are for the five-qubit and seven-qubit cases, respectively. The dashed lines are the ideal results and the solid lines are the results in consideration of the imperfect EPR state preparation, Bell measurement and decoherence. \textbf{d, e}, The left panel shows the evolution of average OTOC under various mismatch strength $\Delta J$ in the $\sigma_+^i \sigma_-^{i+1} + \sigma_-^i \sigma_+^{i+1}$ interaction term, and the right panel represents the simulated OTOC dynamics (decoherence included) with the change of the mismatch. \textbf{d} and \textbf{e} are for the five-qubit and seven-qubit cases, respectively.}
\label{fig:FigS1}
\end{figure*}

Here we present a set of parameters in Table~\ref{Table:TableS1} which is used in our experiment based on the nonintegrable Hamiltonian containing $\sigma_z^i$, $\sigma_z^i \sigma_z^{i+1}$ and $\sigma_x^i$ terms. The corresponding experimental results are depicted in Fig.1 in the main text and the device parameters can be found in Table~\ref{Table:TableS3} and Table~\ref{Table:TableS4}. The resonant Hamiltonian containing $\sigma_z^i \sigma_z^{i+1}$ and $\sigma_x^i$ terms can be realized with $\Delta_i=0$.

\subsection{OTOC dynamics in $(XX+YY)$-type coupling scheme}

Apart from the realization of the $ZZ$-type coupling, the $(XX+YY)$-type coupling can also be generated in our system. We also simulate the integrable Hamiltonian based on the $\sigma_+^i \sigma_-^{i+1} + \sigma_-^i \sigma_+^{i+1}$ interaction to study the scrambling dynamics. Our quantum processor can naturally generate the Hamiltonian:
\begin{equation}\label{Eq1-2}
\begin{split}
H&=-\sum_{i=1}^{N} \frac{\omega_i}{2} \sigma_z^{i} + \sum_{i=1}^{N-1} J_{i,i+1}(\sigma_+^i \sigma_-^{i+1} + \sigma_-^i \sigma_+^{i+1}),
\end{split}
\end{equation}
where $\omega_i$ is the bare frequency of the qubits and $J_{i,i+1}$ is the effective coupling strength between each neighbouring qubit pair. By tuning the qubit system into resonance and simultaneously opening the coupler \cite{Yan_2018,Li_2020} to turn on the coupling, in the interaction picture of the qubit frequency, the Hamiltonian can be further expressed as:
\begin{equation}\label{Eq1-3}
\begin{split}
H&=\sum_{n=1}^{N-1} \frac{J_{i,i+1}}{2} (\sigma_x^{i} \sigma_x^{i+1} + \sigma_y^{i} \sigma_y^{i+1}),
\end{split}
\end{equation}
which is a special case of the $XXZ$ model with $J_z=0$. In our experiment, to realize this interaction, we tune the qubit system into the resonant frequencies and simultaneously open the coupler to turn on the coupling. Taking seven-qubit case as an example, the qubits in the subsystem $Q_1-Q_3$ are tuned into resonance during the OTOC dynamics with a positive coupling strength, while inversely, the subsystem $Q_4-Q_6$ is in resonance with a negative coupling strength. The corresponding pulse sequence is depicted in Fig.~\ref{fig:FigS1}\textbf{a}. Then we can take advantage of the OTOC measurement to explore the dynamics during the evolution.

\subsubsection{Experimental parameters}

Similarly, we here first present the experimental parameters used in $(XX+YY)$-type coupling scheme as shown in Table~\ref{Table:TableS2}.

\begin{table}[ht]
\caption{Hamiltonian parameters in $(XX+YY)$-type couplings scheme. $J_{i,i+1}$ is the $\sigma_+^i \sigma_-^{i+1} + \sigma_-^i \sigma_+^{i+1}$ interaction between the nearest neighbor qubits in the subsystem in OTOC dynamics.}
\begin{threeparttable}
\begin{tabular}{cp{1cm}<{\centering}p{0.9cm}<{\centering}p{0.9cm}<{\centering}p{0.9cm}<{\centering}p{0.9cm}<{\centering}p{0.9cm}<{\centering}p{0.9cm}<{\centering}p{0.9cm}<{\centering}}
\\[-2pt]
&\multicolumn{4}{c}{Five Qubit OTOC Experiment} \tabularnewline
\\[-7pt]
\hline
\hline
&{$Q_{1-2}$} &{$Q_{2-3}$} &{$Q_{3-4}$} &{$Q_{4-5}$} &{$Q_{5-6}$} &{$Q_{6-7}$} \tabularnewline
\hline
$J_{i,i+1}/2\pi$ (MHz) &{$3.73$} &{$0$} &{$-3.73$} &{$0$} &{$\sim$} &{$\sim$}  \tabularnewline
\hline

\\[-5pt]
&\multicolumn{4}{c}{Seven Qubit OTOC Experiment} \tabularnewline
\\[-7pt]
\hline
\hline
&{$Q_{1-2}$} &{$Q_{2-3}$} &{$Q_{3-4}$} &{$Q_{4-5}$} &{$Q_{5-6}$} &{$Q_{6-7}$} \tabularnewline
\hline
$J_{i,i+1}/2\pi$ (MHz) &{$3.7$} &{$3.7$} &{$0$} &{$-3.7$} &{$-3.7$} &{$0$}  \tabularnewline
\hline
\end{tabular} \vspace{0pt}
\label{Table:TableS2}
\end{threeparttable}
\end{table}

\subsubsection{OTOC dynamics and parameter mismatch}

We first verify our control ability via characterizing the average OTOC and the teleportation fidelity with the same measurement method clarified in the main text. The experimental results are shown in Fig.~\ref{fig:FigS1}\textbf{b, c} with the simulation. We can find that the average OTOC and the teleportation fidelity oscillate during the evolution time under the integrable $(XX+YY)$-type Hamiltonian. Besides, we also measure the effect of Hamiltonian mismatch with the change of the effective coupling strength $J_{i,i+1}$. The corresponding measurement results are plotted in Fig.~\ref{fig:FigS1}\textbf{d, e}. Again, with the increase of the mismatch, the average OTOC shows a trend of decline, indicating a deliberate error-induced decay. 

\section{Experimental Setup}

\subsection{Measurement Setup}

\begin{figure*}[hbt]
\includegraphics{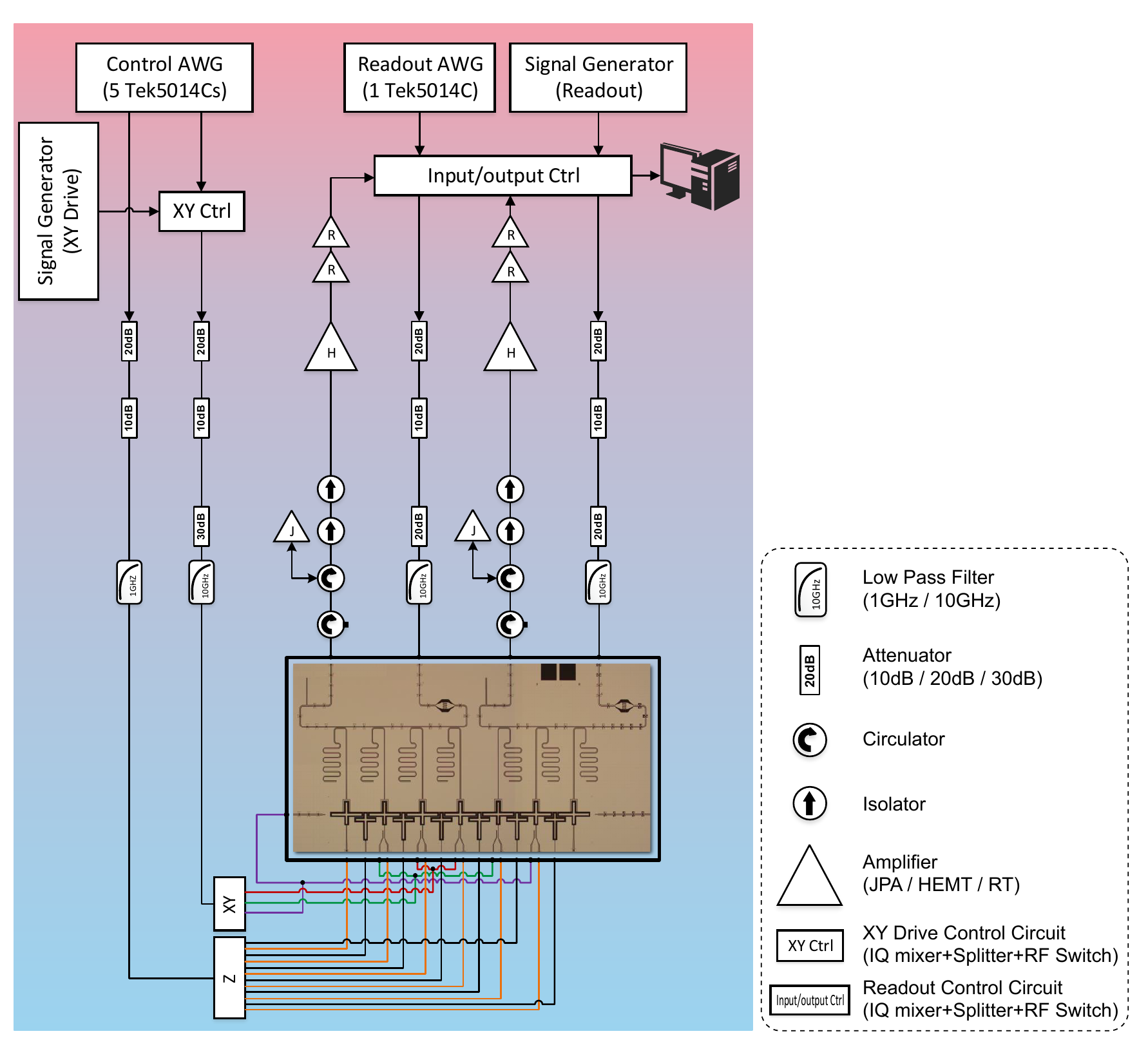}
\caption{\textbf{Measurement setup.} 
The schematic measurement circuit for OTOC experiment.}
\label{fig:FigS2}
\end{figure*}

\begin{table*}[ht]
\caption{Coupler parameters for quantum processor. $\omega_{c,opt}\, (c=1 \sim 6)$ are the resonant frequencies at the sweet spot. $g_{jk}\, (k=1,\, 2)$ represent the direct coupling strength between the coupler and its neighbouring qubit. $g_{jd}$ represents the direct coupling between the neighbouring qubits (`$j$' denotes the order number of the couplers and `$k$' denotes the left or right neighbouring qubit to the coupler with $k=1$ or $k=2$).}
\begin{tabular}{cp{2.4cm}<{\centering}p{2.4cm}<{\centering}p{2.4cm}<{\centering}p{2.4cm}<{\centering}p{2.4cm}<{\centering}p{2.4cm}<{\centering}p{2.4cm}<{\centering}}
\hline
\hline
&{$C_1$} &{$C_2$} &{$C_3$} &{$C_4$} &{$C_5$} &{$C_6$} \tabularnewline
\hline
$\omega_{c,opt}/2\pi$ (GHz) &{$7.752$} &{$7.808$} &{$7.838$} &{$7.798$} &{$7.822$} &{$7.785$} \tabularnewline
$g_{j1}/2\pi$ (MHz) &{$63$} &{$63$} &{$63$} &{$63$} &{$63$} &{$63$} \tabularnewline
$g_{j2}/2\pi$ (MHz) &{$63$} &{$63$} &{$63$} &{$63$} &{$63$} &{$63$} \tabularnewline
$g_{jd}/2\pi$ (MHz) &{$5.17$} &{$5.55$} &{$5.36$} &{$4.92$} &{$5.07$} &{$\sim$} \tabularnewline
\hline
\end{tabular} \vspace{-6pt}
\label{Table:TableS3}
\end{table*}

\begin{table*}[ht]
\caption{Qubit parameters for seven qubit OTOC experiment. $\omega_{i,idle} \, (i=1 \sim 7)$ are the idle frequencies used for preparing EPR pairs and performing Bell measurement. $\omega_{i,ZZ,OTOC}\, (i=1 \sim 7)$ are the frequencies used for OTOC evolution in the $ZZ$-type coupling regime while $\omega_{i,XX+YY,OTOC}\, (i=1 \sim 7)$ are the frequencies used for OTOC evolution in the $(XX+YY)$-type coupling regime. $\alpha_i \, (i=1 \sim 7)$ are the anharmonicities of each qubit. $T_{1,ZZ,OTOC}$, $T_{2,ZZ,OTOC}$ and $T_{2E,ZZ,OTOC}$ are the corresponding energy relaxation time, Ramsey dephasing time and echoed dephasing time of the qubits measured at the frequencies used for OTOC evolution in the $ZZ$-type coupling scheme, which are used as the corresponding parameters in the numerical simulations. $F_{gg,idle}$ and $F_{ee,idle}$ are fidelities which are detected by measuring the qubits in $\ket{g}$ ($\ket{e}$).}
\begin{threeparttable}
\begin{tabular}{cp{2cm}<{\centering}p{1.9cm}<{\centering}p{1.9cm}<{\centering}p{1.9cm}<{\centering}p{1.9cm}<{\centering}p{1.9cm}<{\centering}p{1.9cm}<{\centering}p{1.9cm}<{\centering}}
\\[-7pt]
&\multicolumn{6}{c}{Five Qubit OTOC Experiment} \tabularnewline
\\[-7pt]
\hline
\hline
&{$Q_1$} &{$Q_2$} &{$Q_3$} &{$Q_4$} &{$Q_5$} &{$Q_6$} &{$Q_7$} \tabularnewline
\hline
$\omega_{i,idle}/2\pi$ (GHz) &{$4.220$} &{$4.425$} &{$4.365$} &{$4.488$} &{$4.547$} &{$\sim$} &{$\sim$} \tabularnewline
$\omega_{i,ZZ,OTOC}/2\pi$ (GHz) &{$4.220$} &{$4.425$} &{$4.365$} &{$4.488$} &{$4.547$} &{$\sim$} &{$\sim$} \tabularnewline
$\omega_{i,XX+YY,OTOC}/2\pi$ (GHz) &{$4.228$} &{$4.228$} &{$4.370$} &{$4.370$} &{$4.547$} &{$\sim$} &{$\sim$} \tabularnewline
$\alpha_i/2\pi$ (MHz) &{$-220$} &{$-218$} &{$-218$} &{$-213$} &{$-222$} &{$\sim$} &{$\sim$} \tabularnewline
$T_{1,ZZ,OTOC}$ ($\mu$s) &{$22.9$} &{$22.5$} &{$22.0$} &{$16.4$} &{$19.9$} &{$\sim$} &{$\sim$} \tabularnewline
$T_{2,ZZ,OTOC}$ ($\mu$s) &{$3.0$} &{$6.7$} &{$6.8$} &{$4.2$} &{$7.9$} &{$\sim$} &{$\sim$} \tabularnewline
$T_{2E,ZZ,OTOC}$ ($\mu$s) &{$8.5$} &{$12.5$} &{$11.5$} &{$10.2$} &{$17.0$} &{$\sim$} &{$\sim$} \tabularnewline
$F_{gg,idle}$ (\%) &{$92.7$} &{$93.5$} &{$94.0$} &{$93.5$} &{$93.2$} &{$\sim$} &{$\sim$} \tabularnewline
$F_{ee,idle}$ (\%) &{$89.3$} &{$87.5$} &{$89.0$} &{$86.1$} &{$85.0$} &{$\sim$} &{$\sim$} \tabularnewline
\hline

\\[-2pt]
&\multicolumn{6}{c}{Seven Qubit OTOC Experiment} \tabularnewline
\\[-7pt]
\hline
\hline
&{$Q_1$} &{$Q_2$} &{$Q_3$} &{$Q_4$} &{$Q_5$} &{$Q_6$} &{$Q_7$} \tabularnewline
\hline
$\omega_{i,idle}/2\pi$ (GHz) &{$4.219$} &{$4.424$} &{$4.238$} &{$4.490$} &{$4.544$} &{$4.362$} &{$4.519$} \tabularnewline
$\omega_{i,ZZ,OTOC}/2\pi$ (GHz) &{$4.219$} &{$4.424$} &{$4.238$} &{$4.490$} &{$4.544$} &{$4.362$} &{$4.519$} \tabularnewline
$\omega_{i,XX+YY,OTOC}/2\pi$ (GHz) &{$4.228$} &{$4.228$} &{$4.228$} &{$4.350$} &{$4.350$} &{$4.350$} &{$4.519$} \tabularnewline
$T_{1,ZZ,OTOC}$ ($\mu$s) &{$22.9$} &{$22.5$} &{$23.9$} &{$16.4$} &{$19.9$} &{$23.3$} &{$24.8$} \tabularnewline
$T_{2,ZZ,OTOC}$ ($\mu$s) &{$2.6$} &{$6.4$} &{$2.4$} &{$3.6$} &{$7.7$} &{$5.0$} &{$11.7$} \tabularnewline
$T_{2E,ZZ,OTOC}$ ($\mu$s) &{$7.9$} &{$12.1$} &{$8.6$} &{$9.9$} &{$16.4$} &{$17.4$} &{$21.9$} \tabularnewline
$F_{gg,idle}$ (\%) &{$94.0$} &{$93.4$} &{$94.1$} &{$93.4$} &{$92.1$} &{$91.5$} &{$91.7$} \tabularnewline
$F_{ee,idle}$ (\%) &{$86.2$} &{$87.1$} &{$90.2$} &{$85.2$} &{$86.3$} &{$87.6$} &{$87.5$} \tabularnewline
\hline
\end{tabular} \vspace{0pt}
\label{Table:TableS4}
\end{threeparttable}
\end{table*}

The quantum processor is mounted in an aluminium sample holder at a base temperature of 10 mK in a dilution refrigerator, protected with a magnetic shielding and an infrared shielding. The experimental setup and the measurement circuit are depicted in Fig.~\ref{fig:FigS2} with the simplified circuit handling to emphasize the major part. The detailed measurement circuitry can be found in Ref. \cite{Cai_2021}. 

Each of the qubits (except for $Q_7$) and the couplers have an individual $Z$-lines to tune the frequency, while the $XY$-lines are combined with cryogenic splitters between the pairs of $Q_1$ and $Q_6$, $Q_2$ and $Q_5$, $Q_3$ and $Q_4$. $Q_7$ is controlled via the microwave crosstalk from the $XY$ drive line of $Q_6$. To fully control the qubits and the couplers, we use six Arbitrary Waveform Generators (AWGs) (Tek5014C), two signal generators and two Alazard digitizer cards (ATS9870) to generate control pulse, adjust flux and perform readout, with delicate synchronization to guarantee a stable phase in quantum circuit implementation. In addition, a Josephson junction parametric amplifier (JPA), pumped and biased by another signal generator and a voltage source, is used with a gain of more than $20$ dB and a bandwidth of about $300$ MHz \cite{kamal2009signal,hatridge2011dispersive,Roy_2015}, followed by a high-electron mobility transistor amplifier at $4$ K and two room-temperature amplifiers for each of the two readout channels, allowing for a high-fidelity simultaneous single-shot readout for all the qubits.

\subsection{Device Parameters}

The parameters of our superconducting quantum processor are listed in Table~\ref{Table:TableS3} and Table~\ref{Table:TableS4}. Table~\ref{Table:TableS3} mainly presents the frequency and the coupling strength for the couplers. Here the qubit-coupler coupling $g_{jk} \, (k=1, \, 2)$ is measured via the qubit-coupler resonant oscillation \cite{Li_2020}. We can extract the exact coupling strength by probing the effective coupling at the resonance position. In addtion, the qubit-qubit coupling $g_{jd}$ can be extracted with the formula $J=\frac{g_{j1} g_{j2}}{2}(\frac{1}{\Delta_{j1}}+\frac{1}{\Delta_{j2}}-\frac{1}{\Sigma_{j1}}-\frac{1}{\Sigma_{j2}})+g_{jd}$ where $\Delta_{jk} \, (k=1, \, 2)$ is the frequency detuning between the $j$th coupler and the two neighbouring qubits , $\Sigma_{jk} \, (k=1, \, 2)$ denotes the frequency summation between each coupler and its neighbouring qubit \cite{Yan_2018}. Given the extracted $g_{jk} \, (k=1, \, 2)$ and the effective coupling strength $J$ at the resonance position, together with the frequencies of the qubits and the couplers, we can calculate the qubit-qubit direct coupling strength as $g_{jd}=J-\frac{g_{j1} g_{j2}}{2}(\frac{1}{\Delta_{j1}}+\frac{1}{\Delta_{j2}}-\frac{1}{\Sigma_{j1}}-\frac{1}{\Sigma_{j2}})$.

Table~\ref{Table:TableS4} shows the detailed parameters for each qubit in five-qubit and seven-qubit cases. According to the different coupling schemes ($ZZ$-type and $(XX+YY)$-type), the corresponding frequencies of qubits in OTOC dynamics are different. We also characterize the energy relaxtion time, Ramsey dephasing time and echoed dephasing time during the OTOC evolution which are used to further simulate the scrambling dynamics in the main text.

\section{Experimental Techniques for Calibration}
The basic system calibration includes detection and suppression of $XYZ$-line crosstalk, calibration of measurement crosstalk, calibration for single-qubit and two-qubit gates and preparation of Bell state.

\subsection{Detection and suppression of $XYZ$-line crosstalk}

\begin{figure}[bt]
\includegraphics{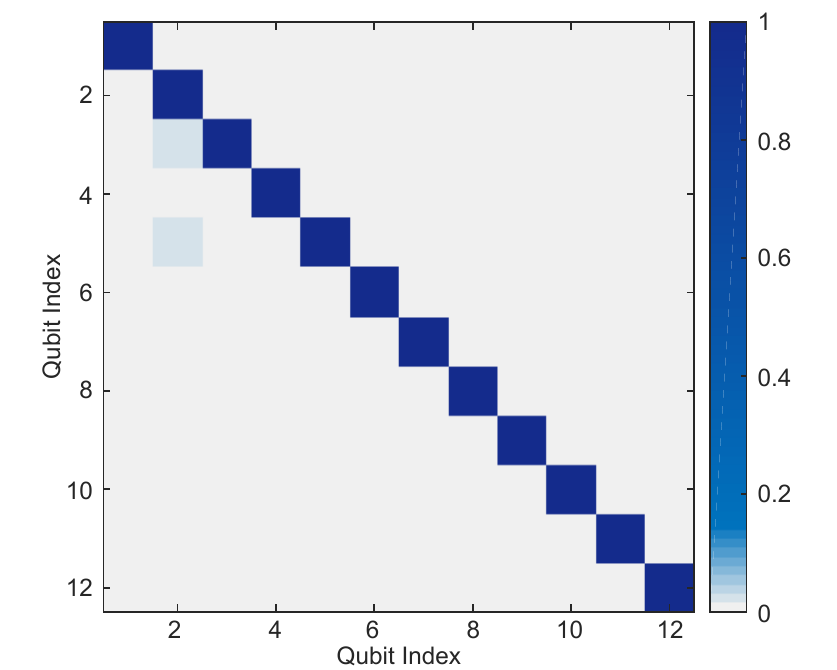}
\caption{\textbf{$Z$-line crosstalk.} 
The $Z$-line crosstalk via measuring the qubit frequency response in order from the qubits $Q_1 - Q_6$ and the couplers $C_1 - C_6$. The $Z$-crosstalk correction matrix can be acquired through the inverse of the crosstalk matrix.}
\label{fig:FigS3}
\end{figure}

\begin{figure}[hbt]
\includegraphics{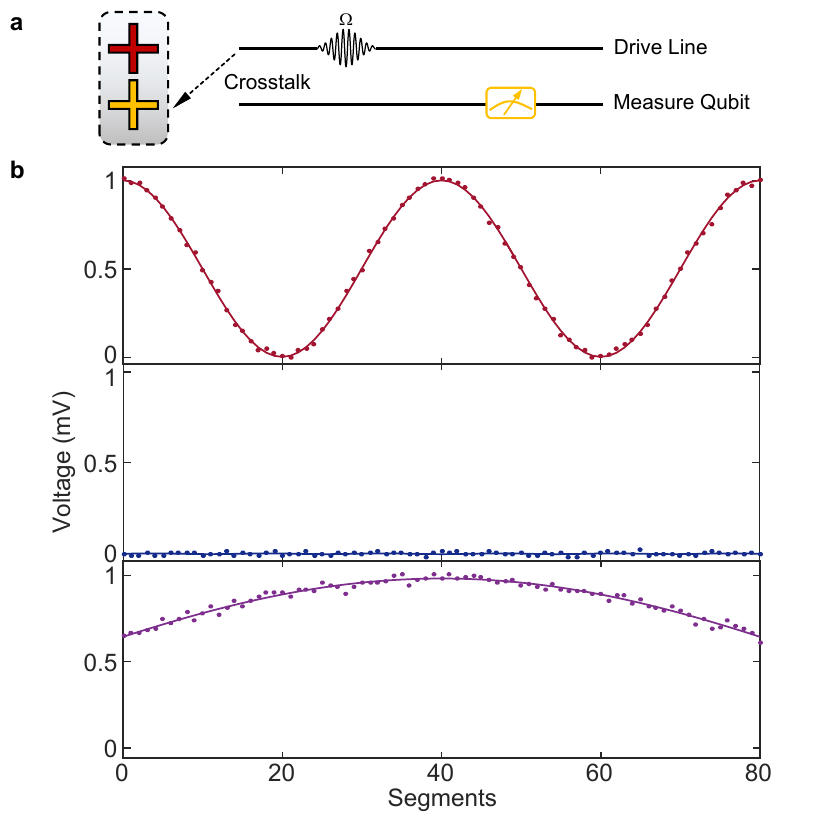}
\caption{\textbf{$XY$-line crosstalk.} 
\textbf{a}, The pulse sequence for detecting $XY$-line crosstalk. The microwave pulse amplitude $\Omega$ is changed to acquire the Rabi oscillation pattern. \textbf{b}, Dectection of the $XY$-line crosstalk. Upper panel: the measured qubit (detector) can be driven with its own frequency but through other $XY$ lines. Middle panel: the measured qubit remains in the ground state if the drive frequency is away from the qubit frequency. Bottom panel: The measured qubit is indeliberately excited once the fequency detuning between the measured qubit and the driven qubit is placed in certain resonance regions.}
\label{fig:FigS4}
\end{figure}

\begin{figure*}[hbt]
\includegraphics{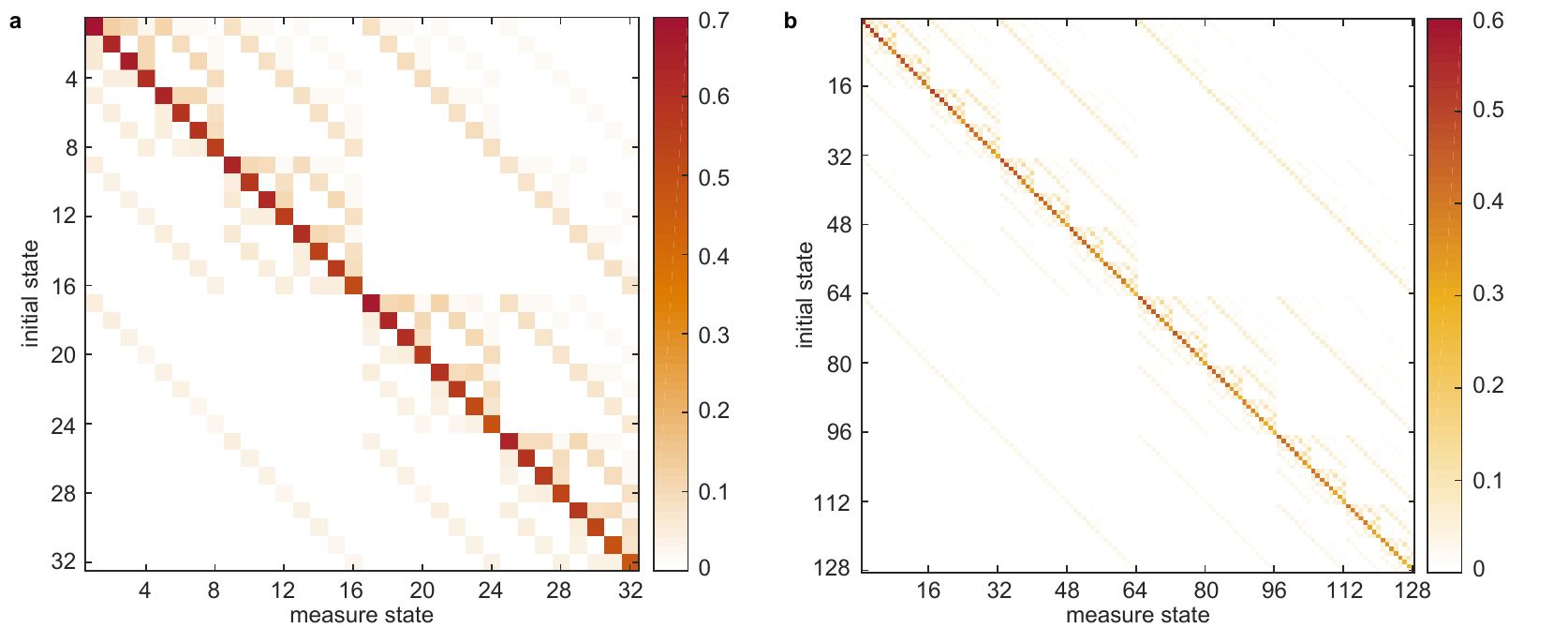}
\caption{\textbf{Measurement crosstalk.} 
\textbf{a, b}, The measurement crosstalk matrix in the five qubit and seven-qubit cases. The $y$-axis represents the initial prepared states, arranged in the order from $|ggggg \rangle$ to $|eeeee \rangle$ (marked as $1-32$) for five-qubit case and $|ggggggg \rangle$ to $|eeeeeee \rangle$ (marked as $1-128$) for seven-qubit case, while the $x$-axis represents the actual measured states in the same order.}
\label{fig:FigS5}
\end{figure*}

To fully control the Hamiltonian in our tunable superconducting qubit system, we need to consider the inevitable $XY$-line crosstalk and $Z$-line crosstalk. $Z$-line crosstalk attributes to the uncontrollable return path of the dc current on each flux line, leading to the presence of the $Z$-line to $Z$-line crosstalk \cite{reed2013entanglement}. In our quantum simulation procedure, this crosstalk may result in an inaccurate control of the qubit and the coupler frequency. The corresponding measurement of the $Z$-line crosstalk is shown in Fig.~\ref{fig:FigS3}. Clearly, the crosstalk seems to be small except for some flux-line pairs; however, this crosstalk still cannot be ignored in a real experimental process. We perform the corretion of the $Z$-line crosstalk via the orthogonalization of the flux lines as shown in our previous work in Ref. \cite{Cai_2021}. 

$XY$-line crosstalk is troublesome in a quantum system especially in a large-scale quantum chip \cite{PhysRevX.11.021010,arute2019quantum,zhu2021quantum}. In fact, each qubit could suffer from the microwave control pulse applied on other arbitraty transmon qubits owing to the always-on capacitive coupling. When the frequency detuning of any two qubits in the quantum chip is close to some special regions, such as $\ket{01}$ and $\ket{10}$ resonant position, $\ket{11}$ and $\ket{02} \, (\ket{20})$ resonant position, two-photon excitation position and two-level systems (TLSs) resonant position, an undesired excitation will occur which ruin the prepared states or operations. In addition, in our experimental setup, the qubit pairs, $Q_1$ and $Q_6$, $Q_2$ and $Q_5$, $Q_3$ and $Q_4$ share one $XY$-line, respectively, and this will enhance the crosstalk. Therefore, the detection and suppresion of $XY$-line crosstalk is essential. Of course, the ideal way to eliminate this $XY$-line crosstalk can be realized through chip design and circuit line optimization. Nevertheless, suppression of the crosstalk can also be achieved according to the experimental situation. For instance, we can use two commonly used suppression methods: frequency arrangement and $XY$-line correction matrix \cite{PhysRevX.11.021010}.

We first clarify the detection of the $XY$-line crosstalk in our experiment which borrows the basic idea of the power Rabi calibration for qubit, that is, the measurement of Rabi oscillation. The detection pulse sequence can be found in Fig.~\ref{fig:FigS4}\textbf{a}. Here, we use the commonly adopted power Rabi sequence with a variation of the microwave amplitude. The only difference is the measured qubit can be different from the driven qubit. For example, if we want to detect the crosstalk from the $Q_2$ $XY$-line, then we can choose $Q_1$ as the measured qubit and the drive is applied on the $Q_2$ $XY$-line. The reason for choosing Rabi oscillation to detect the potential $XY$-line crosstalk is to accurately observe the crosstalk especially some subtle ones. These subtle crosstalks may be indistinguishable in a single power detection but can show oscillation changes in a Rabi oscillation measurement, see Fig.~\ref{fig:FigS4}\textbf{b}. Considering our real experimental requirements and conditions, we suppress the potential $XY$-line crosstalk via delicate arrangement of the qubit frequencies. In our experiment and the simulation in the $ZZ$-type coupling scheme, both the idle frequencies and the evolution frequencies of the qubits are carefully chosen to be in a dispersive regime. Hence, we carefully calculate the frequency detuning between the qubits to make sure that none of the qubit pairs are placed in the specific resonance regions, We verify the crosstalk using the detection pulse sequence with the pulse frequency according to each measured qubit, since the microwave pulses with other driven qubits' frequencies only have a negligible effect of ac stark shift with $\Omega^2/\Delta$ on the measured qubit. 

\subsection{Calibration of measurement crosstalk}

In this tunable superconducting quantum processor, the measurement crosstalk is nonnegligible which will make an impact on single-shot measurement. We can simply estimate the impact of measurement crosstalk with the method similar to the one used for measuring $XY$-line crosstalk shown in Fig.~\ref{fig:FigS4}\textbf{a}. Note that here, we calibrate the measurement crosstalk by preparing the driven qubit in a ground and an excited state respectively while measuring the response from the readout cavity for the neighbouring measured qubit. If no crosstalk happens, then the measurement result from the neighbouring readout cavity should be the same regardless of whether the driven qubit is in the ground or the excited state. Otherwise, the Rabi oscillation pattern will occur if the crosstalk exists, revealing that the neighbouring readout caivty has a coupling channel to the driven qubit. We carefully characterize all the potential crosstalk and find that the measurement crosstalk generally exists only between the adjacent readout cavity and qubit.

Generally, measurement crosstalk can be effectively suppressed via careful chip design or measurement calibration \cite{heinsoo2018rapid}. Here, we follow our previous work in Ref. \cite{Li_2020} to acquire a correction matrix for the measured
crosstalk. Based on the Bayes' rule, we could further distillate our measurement results via the calibration matrix. This correction matrix can not only calibrate the potential measurement crosstalk but also suppress the unwanted thermal population from the thermal excitation states for each qubit. The measurement results of the crosstalk for the five-qubit and seven-qubit cases are depicted in Fig.~\ref{fig:FigS5}\textbf{a, b} respectively. The corresponding correction matrix can be further achieved with the inverse of the crosstalk matrix.

\subsection{Calibration of single-qubit and two-qubit gate}

To conduct the OTOC measurement, Bell state preparation and Bell measurement are required. Here, we will first clarify our gate set calibration in detail and later describe the Bell state preparation. 

\subsubsection{Single-qubit gate}

\begin{figure}[hbt]
\includegraphics{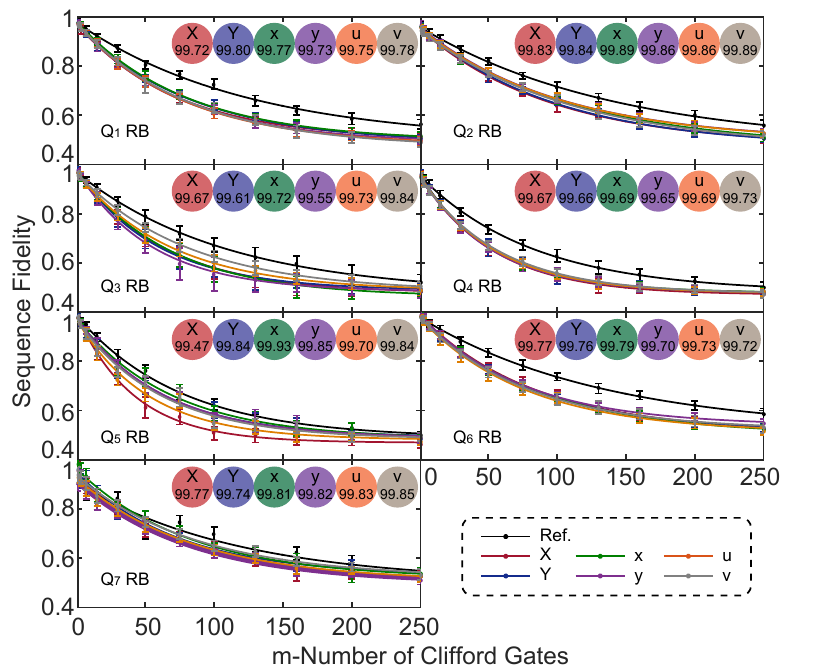}
\caption{\textbf{Single-qubit Randomized Benchmarking.} 
Single qubit interleaved Randomized Benchmarking is implemented with the reference RB and the interleaved RB. We extract the gate infidelity $r_{gate}$ from the reference fidelity $p_{ref}$ and the gate fidelity $p_{gate}$ with the relation $r_{gate}=1-F=\frac{1-p_{gate}/p_{ref}}{2}$. The correponding gate fidelities of six single-qubit rotations for each qubit are depicted in the inset where $X, \, Y, \, x, \, y, \, u, \, v$ represent the single-qubit rotations $R_x(\pi), \, R_y(\pi), \, R_x(\frac{\pi}{2}), \, R_y(\frac{\pi}{2}), \, R_{-x}(\frac{\pi}{2}), \, R_{-y}(\frac{\pi}{2})$ respectively. Further, the average gate fidelity is calculated by averaging the six gate fidelities.}
\label{fig:FigS6}
\end{figure}

To reduce phase error and leakage to the higher energy levels for the typical transmon qubit with a limited anharmonicity, we implement single-qubit gate using the derivative removal adiabatic gate (DRAG) pulse \cite{Motzoi_2009} for all seven qubits. Owing to the limited microwave lines in our dilution refrigerator, $Q_7$ is driven via the $XY$-line of $Q_1$ and $Q_6$. Therefore, we maintain the single-qubit operation time at $60$ ns with the $Z$-only Motzoi pulse for convenience. To ensure the accurate gate operation for single-qubit Clifford group and further calibrate the gate set, we perform interleaved Randomized Benchmarking (RB) to verify the gate fidelity \cite{kelly2015fault}, as depicted in Fig.~\ref{fig:FigS6}. We ramdomly generate 24 rotations in the single-qubit clifford group using microwave pulses only, which can be decomposed into rotations around the $X$ and $Y$ axes. The average single-qubit gate fidelity for each qubit is further calculated via averaging the gate fidelity for six single-qubit rotations shown in the inset of the Fig.~\ref{fig:FigS6}. Note that, during the implementation of the quantum circuit for Bell state preparation and Bell measurement, single qubit rotations in idle position are performed with the neighbouring $ZZ$ interaction off to ensure the isolated single-qubit operations. The measured $ZZ$ interaction is only approximated to be several kHz to tens of kHz. In addition, simultaneous RB is further implemented to ensure the least impact of spectator qubits on the gate operation even if spectators are not in ground state.

\subsubsection{Two-qubit gate}

To generate Bell state for different qubit pairs, two-qubit entangling gates combined with single-qubit rotations are required. In our experiment, we use three types of entangling gates based on the tunable coupler according to the implementation conditions: CZ gate \cite{Li_2020}, iSWAP gate \cite{Li_2020} and parametric iSWAP gate \cite{Han_2020}. 

\begin{figure}[bt]
\includegraphics{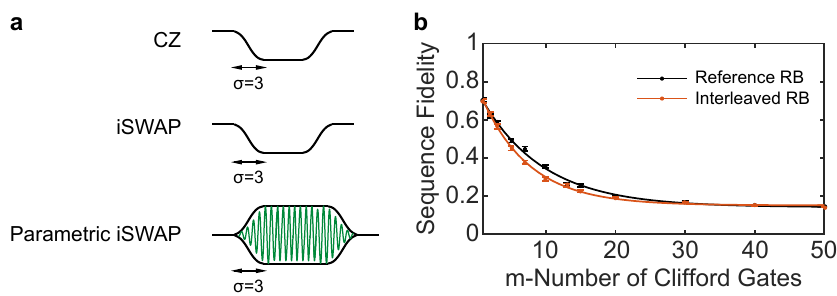}
\caption{\textbf{Two-qubit gates.} 
\textbf{a}, The pulse evelope for the two-qubit entangling gates. \textbf{b}, Two qubit interleaved Randomized Benchmarking. We extract the gate infidelity $r_{gate}$ from the reference fidelity $p_{ref}$ and the gate fidelity $p_{gate}$.}
\label{fig:FigS7}
\end{figure}

The coupler-based CZ gate and iSWAP gate is implemented with simplified pulse shaping. Considering the complexity of the full circuit, the potential crosstalk between qubits and the residual $ZZ$ interaction during the gate operation, we use a simple Gaussian envolop for both the qubit flux pulse and the coupler flux pulse during ascending and descending stage of the gate operation, keeping a balance between gate operation time and gate fidelity, as shown in Fig.~\ref{fig:FigS7}\textbf{a}. Here, a faster gate speed takes less gate time but sacrifices gate fidelity due to the required operation regime where the coupler frequency may approach closely to the qubit frequency. Instead, longer gate time may affect the state fidelity with the consideration of the qubit coherence time. In the five-qubit case, CZ gate time is fixed to be $86$ ns. In the seven-qubit case, CZ gate is implemented in $68$ ns while iSWAP gate is implemented in $67$ ns. As an example, we measure the two-qubit CZ gate fidelity via the two-qubit Randomized Benchmarking, as shown in Fig.~\ref{fig:FigS7}\textbf{b}. The corresponding gate fidelity is around 98\%. 

In the seven-qubit case, for convenience we adopt the parametric iSWAP gate to generate the entanglement between $Q_6$ and $Q_7$ due to the limited RF cables in our refrigerator available for the $Q_7$. Here, we implement parametric iSWAP gate using the parametric modulation pulse applied on the coupler $C_6$. The corresponding pulse shape is depicted in Fig.~\ref{fig:FigS7}\textbf{a}. With the modulation pulse to the coupler, the generated Hamiltonian in the interaction picture can be stated as \cite{reagor2018demonstration}:
\begin{equation}\label{Eq3-1}
\begin{split}
H&=\sum_n J_{6,7,n} e^{i(n\omega_{\phi} - \Delta_{\Omega})t} \ket{10} \bra{01} \\
&+\sum_n \sqrt{2} J_{6,7,n} e^{i(n\omega_{\phi} - (\Delta_{\Omega}+\alpha_7))t} \ket{20} \bra{11} + \cdots,
\end{split}
\end{equation}
where $J_{6,7}$ is the effective coupling strength between $Q_6$ and $Q_7$, $\Delta_{\Omega}=\omega_6-\omega_7$ is the modified frequency detuning due to the modulation drive, $\alpha_i \, (i=6, \, 7)$ are the qubit anharmonicities and $\omega_{\phi}$ is the drive frequency of the modulation pulse. Considering its faster gate speed, here we choose the parametric iSWAP gate to realize the two-qubit entangling gate. Apparently, the parametric iSWAP gate can be implemented via the modulation frequency $n\omega_{\phi}=\Delta_{\Omega}$ with a gate time fixed as $90$ ns.

\subsection{Preparation of Bell state}

\begin{figure*}[hbt]
\includegraphics{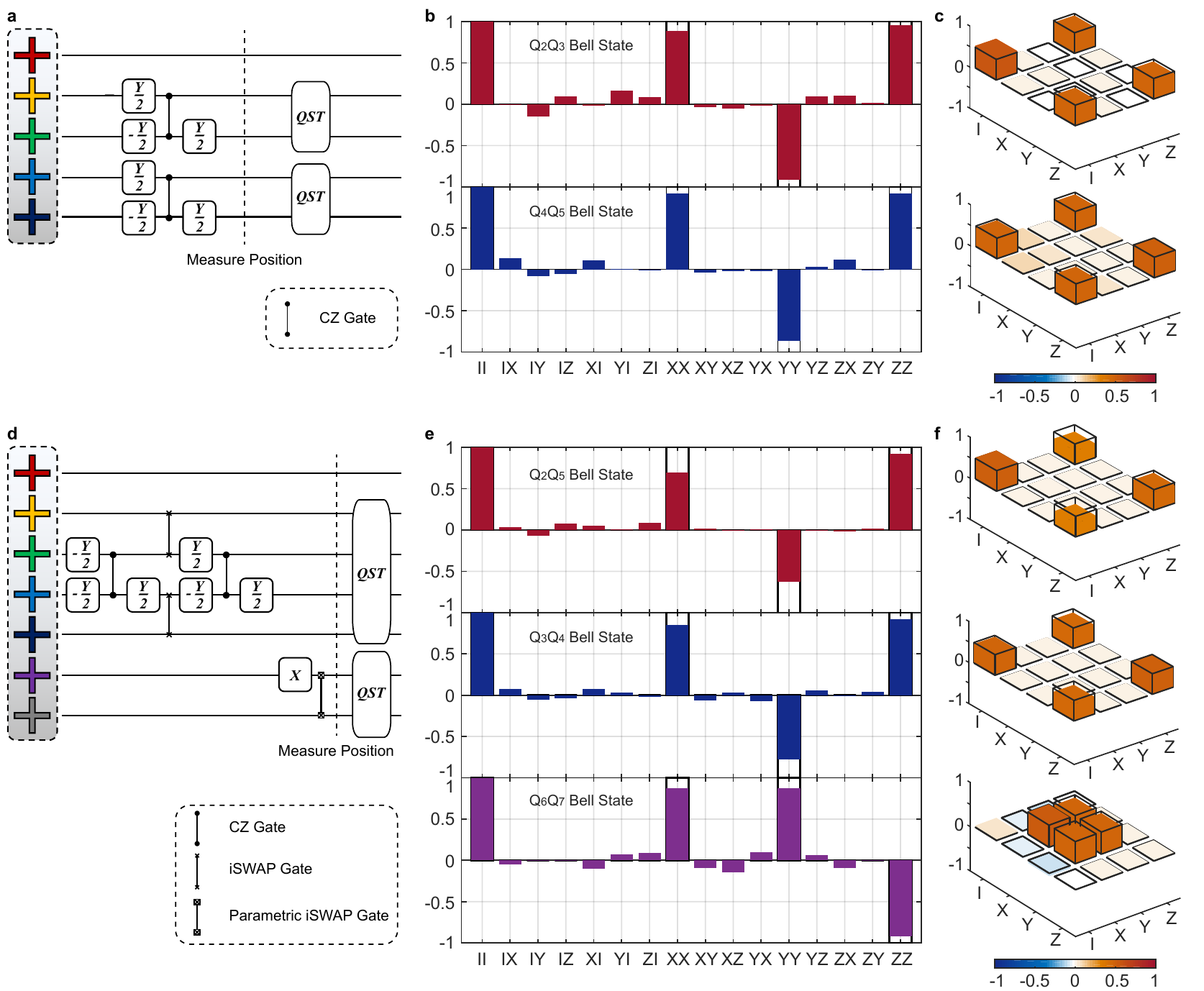}
\caption{\textbf{Preparation of Bell state.} 
\textbf{a}, The pulse sequence for generating two pairs of Bell states in the five-qubit case. \textbf{b, c}, The expectation of Pauli set and the density matrix extracted from the QST measurement for the five-qubit case. \textbf{d}, The pulse sequence for generating three pairs of Bell states in the seven-qubit case. \textbf{e, f}, The expectation of Pauli set and the density matrix extracted from the QST measurement for the seven-qubit case.}
\label{fig:FigS8}
\end{figure*}

In our experiment, we prepare two types of Bell state for the OTOC dynamics: neighbouring Bell state and distant Bell state.

In this part, we take the seven-qubit case as an example since in the five-qubit case preparation of Bell state is similar. Based on the theoretical protocol, three pairs of Bell state are needed to be generated between $Q_3$ and $Q_4$, $Q_2$ and $Q_5$, $Q_6$ and $Q_7$. The target Bell state for our OTOC dynamics is $\frac{1}{\sqrt{2}}(\ket{00}+\ket{11})$ for each EPR pair. However, as mentioned in the main text, we switch the definition of $\ket{0}$ and $\ket{1}$ for $Q_6$ for the purpose of changing $J_{5,6} \sigma_z^5 \sigma_z^6$ to $-J_{5,6} \sigma_z^5 \sigma_z^6$, thus the Bell state for $Q_6$ and $Q_7$ should be correspondingly modified to $\frac{1}{\sqrt{2}}(\ket{01}+\ket{10})$. Among these three EPR pairs, only the $\mathrm{EPR}_{2,5}$ has to be distantly generated between $Q_2$ and $Q_5$ while the other two EPR pairs can be prepared between the two neighbouring qubits. .

Fig.~\ref{fig:FigS8}\textbf{d} gives the pulse sequence for generating Bell state for these three EPR pairs. Starting with EPR state of $\frac{1}{\sqrt{2}}(\ket{00}-\ket{11})$ between $Q_3$ and $Q_4$, followed by two simultaneous iSWAP gates to transfer state from $Q_3$ to $Q_2$, and $Q_4$ to $Q_5$, we implement a final state of $\frac{1}{\sqrt{2}}(\ket{00}+\ket{11})$ between $Q_2$ and $Q_5$ with additional virtual $Z$ gate to each single qubit if necessary. Then, we again generate the Bell state for $Q_3$ and $Q_4$ using the CZ gate, followed by a preparation of Bell state between $Q_6$ and $Q_7$ with the parametric iSWAP gate. Notice that we sequentially prepare the Bell state for the pairs of $\mathrm{EPR}_{6,7}$ and $\mathrm{EPR}_{3,4}$, to prevent the residual impact on the CZ gate between $Q_3$ and $Q_4$ from the parametric modulation drive. The total gate length for the initial Bell state preparation is around $621$ ns containing $4$ ns waiting time between each $XY$ drive pulse and each flux pulse. We further perform quantum state tomography (QST) to measure the experimental density matrix for the three EPR pairs, showing state fidelities of $(80.1 \pm 1.6) \%$ ($Q_2$ and $Q_5$), $(89.5 \pm 0.9) \%$ ($Q_3$ and $Q_4$) and $(91.5 \pm 0.7) \%$ ($Q_6$ and $Q_7$), as shown in Fig.~\ref{fig:FigS8}\textbf{e, f}. Similar to the seven-qubit case, the pulse sequence for generating intial Bell state for the five-qubit case is depicted in Fig.~\ref{fig:FigS8}\textbf{a} with a total gate length around $214$ ns. The measurement results by QST are shown in Fig.~\ref{fig:FigS8}\textbf{b, c} with state fidelities of $(93.4 \pm 0.7) \%$ ($Q_2$ and $Q_3$) and $(91.5 \pm 1.3) \%$ ($Q_4$ and $Q_5$).

\section{Calibration Flow for OTOC dynamics}

In this section, we introduce the calibration procedure on the scrambling dynamics parameters with both the $ZZ$-type scheme and the $(XX+YY)$-type scheme.

\subsection{$ZZ$-type coupling scheme}

We implement the Hamiltonian $H = \sum_{i=1}^{N} \left(\Delta_i \sigma_z^{i} + \Omega_i \sigma_x^{i}\right) + \sum_{i=1}^{N-1} J_{i,i+1} \sigma_z^{i} \sigma_z^{i+1}$ on the qubits $Q_1-Q_3$ for the seven-qubit case and $Q_1-Q_2$ for the five-qubit case, and an opposite Hamiltonian $H_{opp} = \sum_{i=1}^{N} \left(- \Delta_i \sigma_z^{i} - \Omega_i \sigma_x^{i}\right) - \sum_{i=1}^{N-1} J_{i,i+1} \sigma_z^{i} \sigma_z^{i+1}$ on the counterpart ($Q_4-Q_6$ for the seven-qubit case and $Q_3-Q_4$ for the five-qubit case)

\subsubsection{$ZZ$ control}

\begin{figure}[bt]
\includegraphics{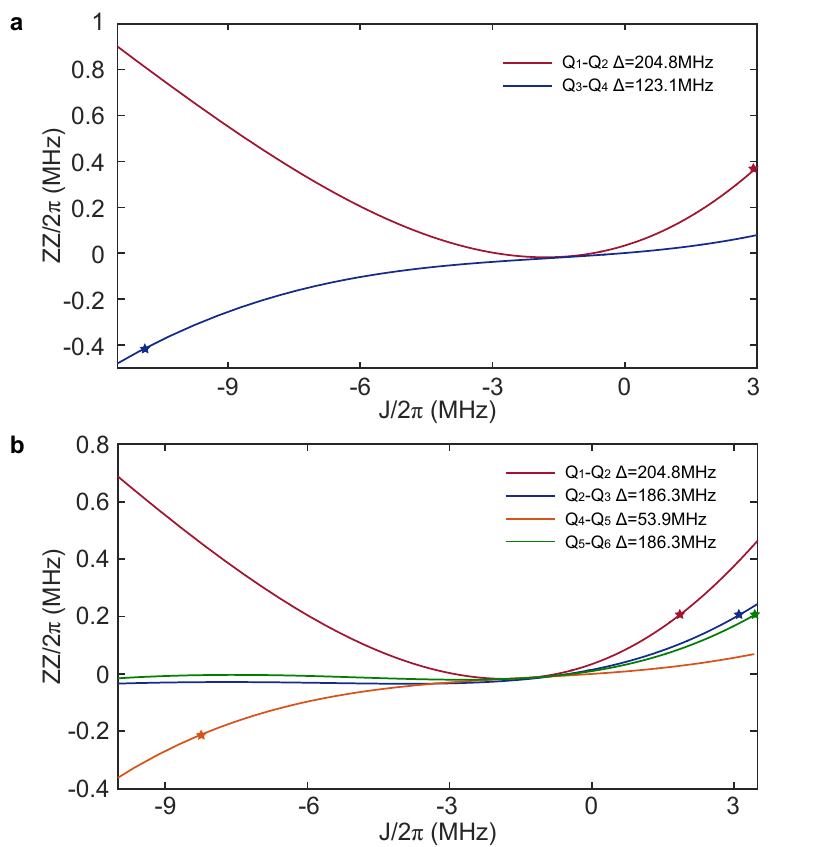}
\caption{\textbf{$ZZ$ control.} 
\textbf{a, b}, The simluated $ZZ$ interaction vs. the effective coupling strength (coupler frequency) between each neighbouring qubit pair in the five-qubit and seven-qubit cases. The star symbol in the figure marks the $ZZ$ interaction chosen in the experiment.}
\label{fig:FigS9}
\end{figure}

We first demonstrate our control of $ZZ$ interaction. We still take the seven-qubit case as an example. In the original scheme, the coupling strength between each neighbouring qubit pair should satisfy $J_{1,2}=J_{2,3}=-J_{4,5}=-J_{5,6}$. However, considering the frequency crowding and the $XY$-line crosstalk, we finally choose the interaction condition as $J_{1,2}=J_{2,3}=-J_{4,5}=J_{5,6}$. Once we switch the definition of $\ket{0}$ and $\ket{1}$ for $Q_6$, the equivalent interaction condition is automatically satisfied. 

Tunable coupler offers us a convinent way to continuously adjust neighbouring qubit-qubit interaction from positive interaction to negative interaction with varying the coupler frequency. We choose the qubit frequency during the scrambling dynamics according to implementation requirement of interaction strength and frequency detuning between the qubits. For example, to generate a relatively large positive $ZZ$ interaction, the frequency detuning between neighbouring qubits should be tuned to be close to the anharmonicity \cite{chu2021coupler}. Fig.~\ref{fig:FigS9}\textbf{b} shows the simulated results for $ZZ$ interaction based on the real experimental parameters listed in Table~\ref{Table:TableS1} as we change the effective coupling strength (coupler frequency) between each neighbouring qubit pair. We can observe that positive $ZZ$ interaction is easier to be achieved with a larger qubit-qubit frequency detuning. Our experimental $ZZ$ interaction is chosen to be around $0.21$ MHz for the seven-qubit case, marked by stars in Fig.~\ref{fig:FigS9}\textbf{b}. Similarly, the simulated $ZZ$ interaction for the five-qubit case is illustrated in Fig.~\ref{fig:FigS9}\textbf{a} with the chosen experimental $ZZ$ interaction about 0.42 MHz, marked by stars in the figure. 

\subsubsection{$X$ control}

\begin{figure}[bt]
\includegraphics{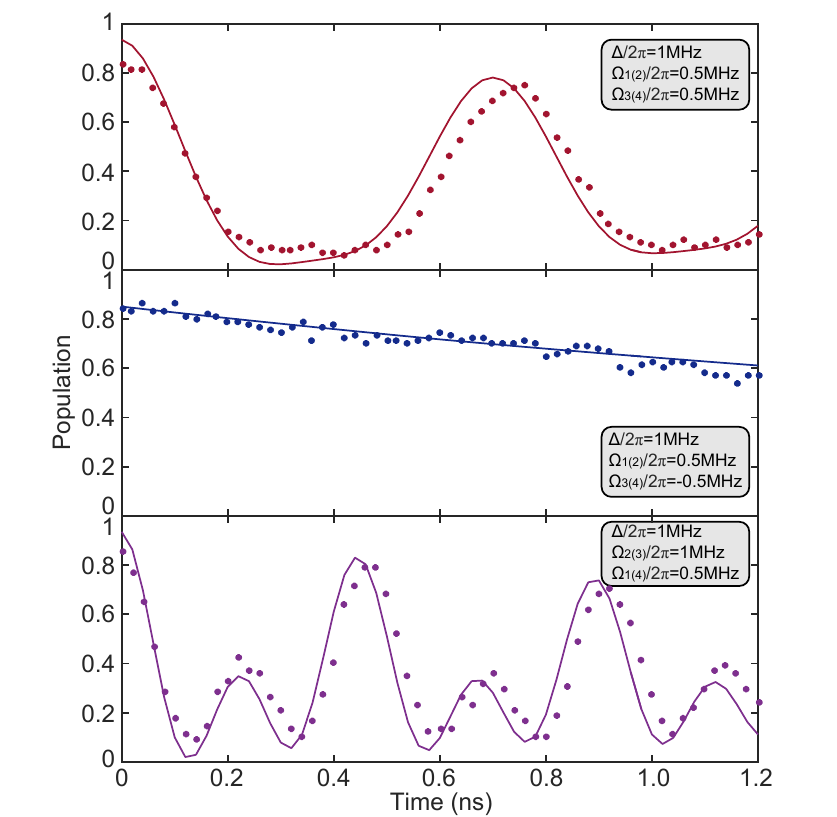}
\caption{\textbf{$X$ control.} 
Verifying the $X$ control by comparing the OTOC measurement with the simulation result. The pulse sequence is similar to that shown in Fig.1 in the main text but with the $ZZ$ coupling off.}
\label{fig:FigS10}
\end{figure}

In the $ZZ$-type coupling scheme, since the qubit frequency remains the same on the idle positions and during the scrambling evolution, thus the driven axis can be fixed according to each qubit frequency. To calibrate the exact pulse amplitude to achieve the requirement $\Omega_1=\Omega_2=\Omega_3=-\Omega_4=-\Omega_5=-\Omega_6$ for the seven-qubit case (similar calibration for the five-qubit case), we implement the Rabi oscillation experiment with varying the microwave pulse amplitude. We use the standard rectangular drive pulse with a gate length of $250$ ns and measure Rabi oscillation to extract the required pulse amplitude.

We verify the dedicate control of the $\sigma_x^i$ term in the whole qubit system by comparing the OTOC measurement with the simulation result. The verification circuit is similar to the formal OTOC experiment circuit shown in Fig.1 in the main text. The only difference in this case is that the $ZZ$ coupling is off during the OTOC dynamics. We find that the measurement of OTOC is actually a fine detector, which can be used to accurately diagnose whether the microwave pulse amplitude and the rotation axis of qubits are correctly calibrated, since the measurement result will show a dramatic change once the pulse amplitude or the rotation axis has a slight deviation from the optimum. Fig.~\ref{fig:FigS10} shows an example of calibrating the $\sigma_x$ term in the five-qubit measurement. It is easy to see that the experimental results of OTOC can be fitted well to the simulation results by optimizing the pulse amplitude and the rotation axis.

\subsubsection{$Z$ control}

To investigate the OTOC dynamics with the nonintergrable Hamiltonian, we should generate $\sigma_z^i$ term with the help of the nonresonant drive. In the interaction picture of the drive frequency, the $\sigma_z^i$ term in the Hamiltonian has the coefficient $\Delta_i=\omega_i-\omega_{d_i} \, (i=1, \, 2, \, \cdots)$ where $\omega_{d_i}$ is the drive frequency on each qubit. In the seven qubit case, to fulfull the requirement of the inverse Hamiltonian, the coefficient should satisfy $\Delta_1=\Delta_2=\Delta_3=-\Delta_4=-\Delta_5=-\Delta_6$, and similarly in the five-qubit case. Nevertheless, we should notice that $\omega_i$ in the definition represents the bare frequency of each qubit which is not the measured qubit frequency in the experiment owing to the existence of the $ZZ$ coupling during the OTOC dynamics \cite{le2021robust}. Here, we take advantage of the $ZZ$ coupling strength to recovery and calibrate the bare frequency from the measured qubit frequency.

The fundamental experimental sequence is the Ramsey pulse sequence \cite{reed2013entanglement,chow2010quantum} which can extract both the qubit dressed frequency and the $ZZ$ interaction. Taking $Q_1-Q_3$ qubit system as an example, the Hamiltonian in the OTOC dynamics without drive could be written as:
\begin{equation}\label{Eq4-1}
\begin{split}
H& = -\frac{\omega_1}{2} \sigma_z^1 - \frac{\omega_2}{2} \sigma_z^2  - \frac{\omega_3}{2} \sigma_z^3 + J_{1,2} \sigma_z^1 \sigma_z^2 + J_{2,3} \sigma_z^2 \sigma_z^3,
\end{split}
\end{equation}
where $J_{1,2}, \, J_{2,3}$ are the $ZZ$ coupling strength between $Q_1$ and $Q_2$, $Q_2$ and $Q_3$. To calculate the bare frequency of $Q_2$, $Q_1$ and $Q_3$ are first prepared in ground state, then the Hamiltonian in the $Q_2$ subspace can be expressed as:
\begin{equation}\label{Eq4-2}
\begin{split}
H_{00}& = \bra{00} H \ket{00} \\
&= -\frac{\omega_1}{2} - \frac{\omega_2}{2} \sigma_z^2  - \frac{\omega_3}{2} + J_{1,2} \sigma_z^2 + J_{2,3} \sigma_z^2 \\
&=-\frac{\omega_2-2J_{1,2}-2J_{2,3}}{2} \sigma_z^2- \frac{\omega_1}{2}- \frac{\omega_3}{2}.
\end{split}
\end{equation}
The measured dressed qubit frequency of $Q_2$ will be $\omega_2^{00}=\omega_2-2J_{1,2}-2J_{2,3}$. Then $Q_1$ is initialized to the excited state while $Q_3$ is still in the ground state. The Hamiltonian in the $Q_2$ subspace can be further expressed as:
\begin{equation}\label{Eq4-3}
\begin{split}
H_{10}& = \bra{10} H \ket{10} \\
&=\frac{\omega_1}{2} - \frac{\omega_2}{2} \sigma_z^2  - \frac{\omega_3}{2} - J_{1,2} \sigma_z^2 + J_{2,3} \sigma_z^2 \\
&=-\frac{\omega_2+2J_{1,2}-2J_{2,3}}{2} \sigma_z^2+ \frac{\omega_1}{2}- \frac{\omega_3}{2}.
\end{split}
\end{equation}
The measured dressed qubit frequency of $Q_2$ in this case will be $\omega_2^{10}=\omega_2+2J_{1,2}-2J_{2,3}$. Combining the measured dressed frequency for $Q_2$ in these two cases, $J_{1,2}$ can be easily extracted. Similarly, $J_{2,3}$ can also be acquired with $Q_1$ prepared in ground state while $Q_3$ prepared in excited state. After that, the bare frequency of $Q_2$ can be calculated as:
\begin{equation}\label{Eq4-4}
\begin{split}
\omega_2 = \omega_2^{00} + 2J_{1,2} + 2J_{2,3}.
\end{split}
\end{equation}
In our real experiment, each qubit bare frequency in the OTOC dynamics is set to be the same as the idle frequency so as to simplify the circuit complexity and calibration flow.

\subsubsection{Phase calibration}

\begin{figure}[bt]
\includegraphics{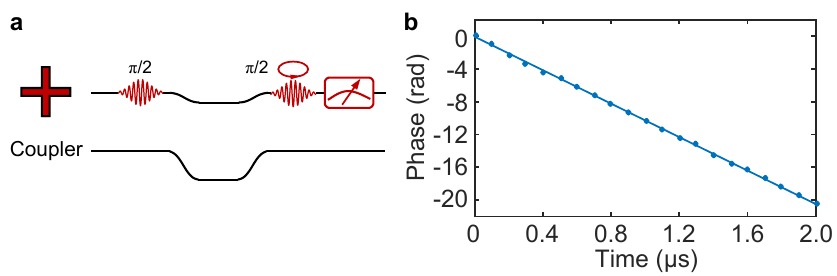}
\caption{\textbf{Phase calibration.} 
\textbf{a}, The pulse sequence for measuring phase accumulation. \textbf{b}, The phase accumulation vs. OTOC evolution time. The linear fitting function is implemented to acquire the phase calibration function.}
\label{fig:FigS11}
\end{figure}

Although in our previous experimental setup, we have claimed that the bare qubit frequencies in OTOC dynamics remain the same to that on the idle position for simplicity. However, owing to the frequency adjustment of the coupler frequencies during the OTOC dynamics, qubit frequencies will still be affected. Therefore, the rotation axes for all qubits should be carefully checked and calibrated to ensure the right axes for implementing the following gate oprations. For the seven-qubit case, the major phase calibration should be accomplished for $Q_3, \, Q_4$ and $Q_7$ prior to the following operations. Similarly, phase calibration should be performed on $Q_2, \, Q_3$ and $Q_5$ in the five-qubit case.

The basic idea for measuring phase accumulation due to the frequency adjustment is the Ramsey experiment. The pulse sequence for phase calibration is depicted in Fig.~\ref{fig:FigS11}\textbf{a}, with the last $\pi/2$ pulse rotating around the axis. We numerically extract the phase accumulation from the Ramsey experiment and calculate the phase-dependent calibration function with the change of the time, as shown in Fig.~\ref{fig:FigS11}\textbf{b} as an example. Notice that the linear fitting function is implemented for phase accumulation as $\phi=kt+b$. Then in the formal OTOC experiment, the rotation axes for gate operations after OTOC dynamics should be modified according to the fitting function of phase calibration.

\subsection{$(XX+YY)$-type coupling scheme}

\subsubsection{$XX+YY$ control}

In this part, we explain the calibration flow for $(XX+YY)$-type coupling scheme in the OTOC experiment, namely the Hamiltonian $H = \sum_{i=1}^{N-1} J_{i,i+1} (\sigma_+^{i} \sigma_-^{i+1} + \sigma_-^{i} \sigma_+^{i+1})$, which can be generated by tuning the qubits into resonance in the interaction picture. In the seven-qubit case, to satisfy the time-inverse Hamiltonian and demonstrate the fully control of the system Hamiltonian, we set the coupling strength to fulfull $J_{1,2}=J_{2,3}=-J_{4,5}=-J_{5,6}$. Similarly, the coupling strength satisfies $J_{1,2}=-J_{3,4}$ in the five-qubit case. Therefore, $XX+YY$ calibration contains two aspects: qubit-qubit resonant position and coupling strength.

\begin{figure}[bt]
\includegraphics{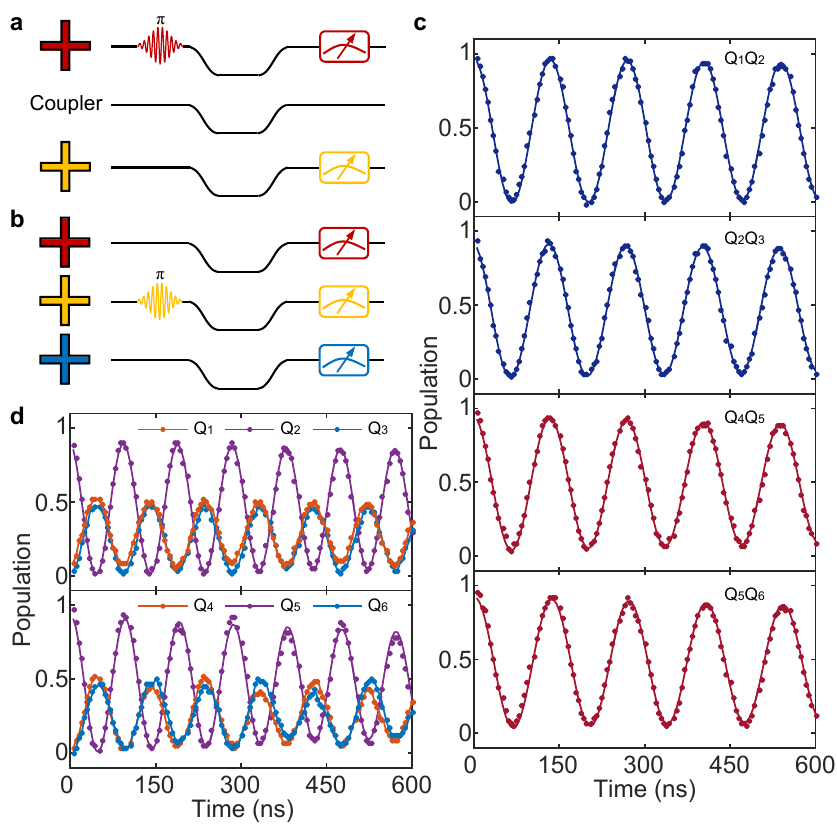}
\caption{\textbf{$XX+YY$ control.} 
\textbf{a}, The pulse sequence for calibrating qubit-qubit resonant coupling. \textbf{b}, The pulse sequence for three qubit resonant oscillations. \textbf{c}, The experimental (dot) and simulation (solid line) results for two qubit resonant coupling. \textbf{d}, The experimental (dot) and simulation (solid line) results for three qubit resonant oscillation which is used to verify the $XX+YY$ calibration.}
\label{fig:FigS12}
\end{figure}

The exact resonant position is calibrated via the qubit-qubit resonant oscillation. The Ramsey measurement is first implemented to make sure all the qubits have been tuned to the positions for the OTOC dynamics. Then, each neighbouring qubit pairs are seprately adjusted to the resonant position with the initial state $\ket{01}$ (or $\ket{10}$). After dedicate adjustment to the qubit flux, we can finally observe the maximum qubit-qubit oscillation pattern at the resonant position for every qubit pair. Notice that, the coupler frequency is simultaneously adjusted with the qubit frequency to ensure the generation of the expected coupling strength. The positive and negative $XX+YY$ interaction could be easily acquired with fine tuning the coupler to different frequencies according to the relation $J=\frac{g_{j1} g_{j2}}{2}(\frac{1}{\Delta_{j1}}+\frac{1}{\Delta_{j2}}-\frac{1}{\Sigma_{j1}}-\frac{1}{\Sigma_{j2}})+g_{jd}$. The experimental calibration results for the seven-qubit case have been plotted in Fig.~\ref{fig:FigS12}\textbf{a} and we can find that the effective coupling strength for all the qubit pairs is set to be the same.  

We finally verify the calibration in the seven qubit case with the simulated three-qubit resonant oscillation. Through simuation, we find that once we bring $Q_1$, $Q_2$ and $Q_3$ into resonance at the same time, the oscillation will feature a finger-print pattern and any deviation from the perfect condition, such as of a slight non-resonance or a mismatched coupling, will modify the regular oscillation pattern. The corresponding experimental results are drawn in Fig.~\ref{fig:FigS12}\textbf{c} combined with the simulation curves, with the pulse sequence shown in Fig.~\ref{fig:FigS12}\textbf{b}. It can be seen that the calibrations of resonant oscillations are well performed among $Q_1-Q_3$ and $Q_4-Q_6$.

\subsubsection{Phase calibration}

In the $(XX+YY)$-type coupling regime, since all the qubits should be accurately tuned for the OTOC dynamics with precisely adjusting the qubit frequencies and the coupler frequencies, the phase calibration is necessary and essential. As we mentioned above, the phase calibration can be realized for every qubit with Ramsey measurement. Notice that, owing to the large frequency tunability of the qubits in the experiment, a range of $0-2\pi$ of the phase accumulation should be considered, and the phase needs to be carefully calibrated by fitting the phase calibration function. The corresponding results are similar to that shown in Fig.~\ref{fig:FigS11} with a same pulse sequence for the measurement. 

\section{Hamiltonian Model and Numerical Simulation}

We verify our accurate control of the superconducting quantum processor by comparing the experimental results with the simulation results based on the Hamiltonian model in the OTOC dynamics. The numerical simulation is performed with Qutip in PYTHON \cite{qutip}. Here, we introduce our simulation metheds and clarify the Hamiltonian models implemented in OTOC dynamics in our experiments. 

\subsection{Hamiltonian models}

According to the experiments demonstrated in our main text and in the extended data, we list all the simulated Hamiltonians for clarity:
\begin{description}
\item[(1)] $H=\sum_{i=1}^{N} \Delta_i \sigma_z^i + \sum_{i=1}^{N-1} J_{i,i+1} \sigma_z^i  \sigma_z^{i+1} + \sum_{i=1}^{N} \Omega_i \sigma_x^i$.
\item[(2)] $H= \sum_{i=1}^{N-1} J_{i,i+1} \sigma_z^i  \sigma_z^{i+1} + \sum_{i=1}^{N}  \Omega_i\sigma_x^i$.
\item[(3)] $H= \sum_{i=1}^{N-1} J_{i,i+1} \sigma_z^i  \sigma_z^{i+1}$.
\item[(4)] $H= \sum_{i=1}^{N-1} \frac{J_{i,i+1}}{2} (\sigma_x^i  \sigma_x^{i+1}+\sigma_y^i  \sigma_y^{i+1})$.
\end{description}
As we mentioned previously, the first three Hamiltonian models represent the $ZZ$-type coupling scheme and the last one represents the $(XX+YY)$-type coupling scheme. These Hamiltonian models have revealed the interesting but different evolution dynamics, as shown in the experiment and simulation results in the main text and the previous sections.

\subsection{Methods of numerical simulation}

In order to realistically simulate the dynamic process of the OTOC evolution based on the experimental conditions, the combined effects of qubits' coherence time, initial Bell state preparation error and Bell measurement error are all taken into account. By including these errors into numerical simulation, we can focus on the difference between the desired Hamiltonian evolution and the one achieved in the experiment.

First, the coherence time is considered by implementing Lindblad master equation based on the experimental coherence time measured in the OTOC dynamics, shown in Table~\ref{Table:TableS4}. The energy relaxation time $T_1$ of all the qubits is included with the Lindblad operator chosen as $a_i \, (i=1\sim7)$ while the dephasing time $T_2$ of all the qubits is considered with the Lindblad operator chosen as $a_i^\dag a_i \, (i=1\sim7)$. The simulated OTOC evolution time and steps are in consistent with the experimental condition. Through our simulation, we find that the presence of the decoherence weakens the average OTOC and influences the noise parameter, thus degrading the signal for quantum scrambling.

Moreover, the initial Bell state preparation is also carefully considered according to the experimental measurement. We use the same pulse sequence as in the actual experiments to accurately characterize the QST of all EPR pairs before the OTOC dynamics. The corresponding measured density matrix is extracted to replace the ideal Bell state in the simulation. We observe that the initial Bell state preparation error will lower down the profile of the average OTOC and the noise parameter, but does not change the overall trend of the evolution. Therefore, the error from the Bell state preparation actually has little impact on the OTOC dynamics.

Finally, the complicated Bell measurement circuit in the experiment also brings additional error which may further decline the average OTOC. We separately characterize the gate fidelity of Bell measurement via the QPT in both the five-qubit and the seven-qubit case. Then the simulated results after the Hamiltonian evolution are further modified by the imperfect gates before the ideal Bell measurement, whose $\chi$ matrix is given by \cite{nielsen2002quantum}:
\begin{equation}\label{Eq5-1}
\begin{split}
\rho_{fin} = \sum_{m,n} \chi_{mn} E_{m} \rho_{in} E_{n}^{\dagger},
\end{split}
\end{equation}
where $\rho_{in}, \, \rho_{fin}$ represent the input and output states; the matrices $\{E_n\}$ construct a complete Pauli basis and here we adopt the definition of $\{E_n\} = \{I, \, X, \, Y, \, Z\}^{\otimes 2}$. Note that after dealing with the imperfect channel, the simulation results fit well with our experiments as shown in the main text and the previous sections.
